\documentclass[sn-mathphys-num]{sn-jnl}

\usepackage{graphicx}%
\usepackage{multirow}%
\usepackage{amsmath,amssymb,amsfonts}%
\usepackage{amsthm}%
\usepackage{mathrsfs}%
\usepackage[title]{appendix}%
\usepackage{xcolor}%
\usepackage{textcomp}%
\usepackage{manyfoot}%
\usepackage{booktabs}%
\usepackage{algorithm}%
\usepackage{algorithmicx}%
\usepackage{algpseudocode}%
\usepackage{listings}%

\begin{document}

\title[Article Title]{Fast pulsed photoemission from a double quantum well on a dielectric substrate as a dynamic process of inverse LEED leading to the generation of a charge and current density wave}


\author[1]{\fnm{Peisakhovich Yu.G.}}\email{yugp@rambler.ru}

\author*[1]{\fnm{Shtygashev A.A.}}\email{shtygashev@corp.nstu.ru}
\equalcont{These authors contributed equally to this work.}

\affil*[1]{\orgname{Novosibirsk State Technical University}, \orgaddress{\street{20, K.Max Ave.}, \city{Novosibirsk}, \postcode{630073},  \country{Russia}}}


\abstract{Within the framework of the rigorous quantum theory of the atomic photoeffect and the scenario of photoemission from a crystal as an inverse LEED (Low Energy Electron Diffraction) process, pulsed photoemission from a flat thin-film photoemitter formed by a double quantum well on a dielectric substrate is investigated. The spatiotemporal dependences of the charge and current densities are calculated using the density matrix method. Asymptotic estimates are made in terms of the evolution of wave packets and the concepts of the extreme phase and fastest descent. The possibility of generating charge and current density waves as a result of the population and decay of quasi-stationary states of a doublet of a double quantum well under photoexcitation of electrons in conducting layers from inside this well is shown. For comparison, calculations were made for photoemission from a metal film on a substrate without a heterostructure: the double-well heterostructure increases, stabilizes, and extends the photocurrent pulse. A comparison of the contributions of the outgoing and incoming waves of the inverse LEED problem is made, and the relative smallness of the contribution of the incoming wave is quantitatively demonstrated for the first time.}

\keywords{Pulsed photoemission, inverse low-energy electron diffraction, charge and current density wave, surface heterostructure}



\maketitle

\section{Introduction}\label{sec:sec1A}

It is well known that the electronic energy spectrum of a double quantum well can contain doublets of stationary or quasi-stationary quantum states, and pulsed excitation and slow decay of a nonstationary state formed by superposition and interference of states from a narrow band of the electronic spectrum including a doublet can be accompanied by beats in the spatiotemporal distributions of charge and current densities of electrons whose energies belong to such a narrow band. Beats of this kind often accompany a quantum transient process \cite{bib03}-\cite{bib08} after a single pulse excitation and last during the lifetime of quasi-stationary states, which can be much longer than the period of these beats, provided that the transparency of the potential barriers of the heterostructure is sufficiently low and the inelastic processes for electrons are weak.

In our recent articles \cite{bib01},\cite{bib02}, it was shown that coupled oscillations of a doublet of mixed resonance states should manifest themselves not only in the periodic flow of electron density through the middle barrier between wells inside a double well \cite{bib07}-\cite{bib10}, but at energies of the continuous spectrum of electrons above the vacuum level, these oscillations can also be accompanied by the generation of waves of charge and current densities of electrons escaping into external space through the extreme potential barriers. In this case, the wave functions of the resonant doublets describe the delocalized quasi-stationary states of the transverse scattering problem, so that the partial amplitudes of the transmitted and reflected waves have pole features. The frequency of the generated waves is equal to the difference frequency of the doublet, the wave number is equal to the difference between the wave numbers of the free movement of electrons with resonant energies, and the speed of their propagation is the ratio of these quantities. When moving away from the heterostructure, the wave packets of charge and current densities attenuate and blur quite slowly they can be detected and removed from the system by means of electric and magnetic fields of the corresponding structure.

In \cite{bib01},\cite{bib02}, we considered scenarios in which the population of quasi-stationary states of a doublet of a double quantum well occurs as a result of the incidence of an electron wave packet of a suitable structure \textit{from outside} the well onto such a well. In \cite{bib01}, the generation of charge and current density waves arose as a result of scattering on a double well of an electron Gaussian wave packet prepared from pure quantum mechanical states of a stationary scattering problem. In \cite{bib02} it was assumed that a double-well heterostructure formed by three identical tunnel-transparent potential barriers was applied to the flat surface of a bulk photoemitter, playing the role of only an energy filter for photoelectrons, and a photoelectron pulse incident on it is formed inside the photoemitter as a result of absorption of a light pulse with fairly sharp thresholds; it was shown that the effect of modulation formation of waves of charge and current densities can arise not only after the trailing edge of the photoemission electron pulse has passed through the heterostructure, but also after the leading edge has passed through. To describe these processes, we developed and presented in article \cite{bib02} a version of the method for calculating the spatio-temporal dependences of charge and current densities, suitable for fairly fast pulsed photoemission processes occurring in a time less than the time of complete relaxation to thermodynamic equilibrium, while one-electron Green's functions can be effectively factored, taking out from under the sign of statistical averaging the wave functions of stationary states of the electron, which carry the coordinate dependence. Then the time dependence of the sought quantities is completely determined by the elements of the density matrix, which satisfy the known kinetic equation of quantum optics \cite{bib11}-\cite{bib16}. We obtained solutions to this equation for the cases of abrupt switching on and off of the pump light pulse. In this case, the partial terms of the sums expressing the charge and current densities are also factorized, their coordinate dependence is described by the diagonal and non-diagonal elements of the effective charge and current density operators, which are multiplied by the corresponding time-dependent elements of the density matrix. Sums over states can be calculated; we did this numerically for states from a narrow region of the electronic energy spectrum, including a significant contribution from the doublet of poles of scattering amplitudes on a double quantum well. Interpretation and analytical estimates of these contributions are possible in terms of the evolution of wave packets and the extreme phase method. It is shown in \cite{bib02} that in this case the effect of modulation of charge and current density waves is the result of a combination of two transient processes: 1) turning on or off photoemission excitation during the action of the leading or trailing edge of the light pump pulse (determines the corresponding time dependence of the density matrix) and 2) tunnel population or decay of quasi-stationary states in a double quantum well (the lifetime of which $\tau  \approx \hbar /E''$ is determined by the imaginary parts  $E''$ of the energies of the poles of the scattering amplitudes). To reduce the blurring of the wave oscillations under study, the duration $\Delta t$ of the light pumping fronts should be short compared to the period   $T$ of doublet beats, and the latter should be small relative to the lifetime  $\tau $ of quasistationary states. Also, the smearing  $\gamma$ of stationary state energies due to electron-electron and electron-phonon scattering within the photoemitter and heterostructure must be sufficiently small  $\Delta t \le T \ll \tau  \le {1 \mathord{\left/ {\vphantom {1 \gamma }} \right.  \kern-\nulldelimiterspace} \gamma }$ to ensure quantum coherence. With all this, the duration  ${t_0}$ of the photoexciting laser pulse and, accordingly, the electron pulse incident on the heterostructure may well be long. These conditions can best be satisfied by photoexciting electrons with a laser pulse that has nearly rectangular fronts (in the limit  $\Delta t \to 0$). The creation of laser pulses of almost rectangular shape with fronts of femtosecond and picosecond duration is not a simple technical problem, which is now arousing quite a lot of interest and is being solved in different ways \cite{bib17},\cite{bib22}.

It is interesting to investigate the scenario when the generation of charge and current density waves occurs as a result of the population of a doublet of quasi-stationary states of a double quantum well, which occurs by photoexcitation of electrons directly in the conducting layers \textit{from inside} this well, which itself serves as a thin-film photoemitter. Such a process cannot at all be interpreted within the framework of the scenario described at the beginning of the previous paragraph.

In this article we present the results of applying the technique described in \cite{bib02} to the calculation of photoemission from such a thin-film photoemitter. This required somewhat rethinking and changing the interpretation of the results obtained for the bulk photoemitter. To describe photoemission from a flat thin-film photoemitter formed by a double quantum well, it turned out to be necessary to more strictly apply the concepts and reasoning of the quantum theory of atomic and molecular photoelectric effect \cite{bib23}-\cite{bib25} and the based on them concept of photoemission from a crystal as an inverse LEED process \cite{bib26}-\cite{bib31}. In these theories, the main goal is usually to calculate the time-independent probability and effective cross section of the atomic photoelectric effect, or the average stationary current density of photoemission from crystals. In the algorithm of such calculations, the most important role is played by the ''amous'' \cite{bib29} ''incoming'' electron waves, because when solving the problem of time-reversed scattering, it is necessary to take into account particular solutions in the form of electron waves converging towards the atom or photoemitter in order to ensure the completeness of the basic system of eigenfunctions through which the desired physical quantities are expressed.

From the classical and semi-classical intuitive point of view on photoemission in the spirit of Einstein$'$s theory and the three-step model \cite{bib32},\cite{bib33},\cite{bib13}, it is difficult to imagine the existence of such electron waves of photoexcited states coming from outside, and this algorithm looks somewhat artificial. The calculation part of the work \cite{bib02} corresponds to such intuitive concepts, namely, it was assumed that a photoelectron pulse, described in the basis of the direct problem of electron scattering in the same way as wave packets in work \cite{bib01}, falls on the double-well heterostructure from the side of the photoemitter. In this case, the partial amplitudes of the transmitted and reflected waves contain pole features responsible for the formation of resonant doublets, and the measured photocurrent is proportional only to quadratic combinations of the amplitudes of the waves transmitted through the heterostructure. The energy dependence of the partial amplitudes of waves incident on the heterostructure from the side of the photoemitter is smooth (not polar) in the main approximation, it is determined by normalization, therefore, the matrix element of the electron dipole moment responsible for photoexcitation is usually also considered to be a fairly smooth function of the energy of excited states \cite{bib12},\cite{bib13},\cite{bib26},\cite{bib27},\cite{bib34},\cite{bib35}. 

Within the framework of a more rigorous quantum theory of the atomic photoelectric effect and the scenario of photoemission from a crystal as an inverse LEED process, in accordance with the concepts first set forth in the dissertation of M. Stobbe \cite{bib23},\cite{bib24}, the basis wave functions of excited stationary states of photoelectrons far from the surface of the emitting system should contain components with asymptotics in the form of partial waves coming both from inside the photoemitter and from outside it. Outside the photoemitter, a plane-wave exponential must also be added, describing the wave propagating in the direction of the detector, and its partial amplitude is determined by normalization and weakly depends on energy. The partial amplitudes of these incoming waves, as well as of waves inside the heterostructure, are expressed through it using the boundary conditions of the time-reversed scattering problem, therefore, in the presence of a double-well heterostructure, they acquire pole features similar to the features of the amplitudes of direct electron scattering on this heterostructure with additional complex conjugation. In this case, the matrix elements of the electron dipole moment also acquire the same polar features.

The photoelectron charge and current densities are given by the sums derived in \cite{bib02}, which are now taken over these basis states. Such sums can be calculated again. For sums over states from a narrow band of the electronic energy spectrum recorded by the detector, including contributions from doublets of the poles of the amplitudes of inverse scattering on a double quantum well, the pole contributions will lead to wave spatiotemporal dependences of the charge and current densities in the form of waves leaving the photoemitter, almost the same as in \cite{bib02}. There are two reasons for this. Firstly, under the sign of the sums, the pole factors are now simply transferred from the amplitude multiplier of the partial outgoing wave to the multiplier of the dipole matrix element, and secondly, for the physically relevant observation time after turning on the light pump pulse, the contribution to these sums from the incoming partial basis waves is negligible. In this paper, we show this numerically and justify it in terms of the evolution of wave packets and the extreme phase method.

\section{The charge and current densyties of photoemission. Formulation of the problem, choise of model and calculation method}\label{sec:sec2A}

For sufficiently fast pulsed photoemission processes occurring in a time less than the time of complete relaxation to thermodynamic equilibrium, we can represent the charge  $n(r,t)$ and current  ${\bf{j}}(r,t)$ densities of photoelectrons at a point  $r$ at an instant in time  $t$ by the following expressions \cite{bib02}:
\begin{equation}
\label{eq:math:eq1}
n(t,r) = 2\;Sp\left( {\hat \rho (t)\hat n(r)} \right) = 2\sum\nolimits_{p,p'} {{\rho _{p',p}}(t){n_{p,p'}}(r)},
\end{equation}
\begin{equation}
\label{eq:math:eq2}
{\bf{j}}(t,r) = 2\;Sp\left( {\hat \rho (t){\bf{\hat j}}(r)} \right) = 2\sum\nolimits_{p,p'} {{\rho _{p',p}}(t){{\bf{j}}_{p,p'}}(r)},
\end{equation}
where  ${\rho _{p',p}}(t)$ are the elements of the one-electron density matrix  $\hat \rho (t)$ at time  $t$,
\begin{equation}
\label{eq:math:eq3}
{n_{p,p'}}(r) = e\psi _{p'}^*(r){\psi _p}(r),
\end{equation}
\begin{equation}
\label{eq:math:eq4}
{{\bf{j}}_{p,p'}}(r){\kern 1pt} {\kern 1pt}  = {\kern 1pt} {\kern 1pt} i\frac{{e\hbar }}{{2m}}\left[ {(\nabla \psi _{p'}^*(r)){\psi _p}(r) - \psi _{p'}^*(r)(\nabla {\psi _p}(r))} \right]
\end{equation}
- ''matrix elements''  of charge   $\hat n(r)$ and current ${\bf{\hat j}}(r)$ density operators \cite{bib36} at point  $r$,  ${\psi _p}(r)$ - Schrodinger wave function of an electron in a stationary state  $p$. Here we are interested in the charge and current densities of photoexcited electrons, equal to the sums of \eqref{eq:math:eq1} and \eqref{eq:math:eq2} over states  $p$ and  $p'$ of the continuous spectrum of the scattering problem, belonging to a fairly narrow detectable energy band of photoelectrons above the vacuum level of the photoemitter. This band covers the real parts of the energies of one doublet of the poles of scattering amplitudes on a double quantum well, which are responsible for the formation of long-lived quasi-stationary states and make resonant contributions to the sums \eqref{eq:math:eq1} and \eqref{eq:math:eq2}.

	       Our main goal here is to develop a technique that allows us to calculate the wave-like spatiotemporal dependences of the resonant contributions to the photocurrent, determined by the presence of a double-well heterostructure, therefore, for now we limit ourselves to the simplest quasi-one-dimensional model of such a system, which allows for an exact solution for the wave functions of excited states and allows us to advance quite far in understanding the mechanism of the phenomenon and in implementing the numerical calculations of the processes under consideration.

\begin{figure}[h]
\centering
\includegraphics[width=6 cm]{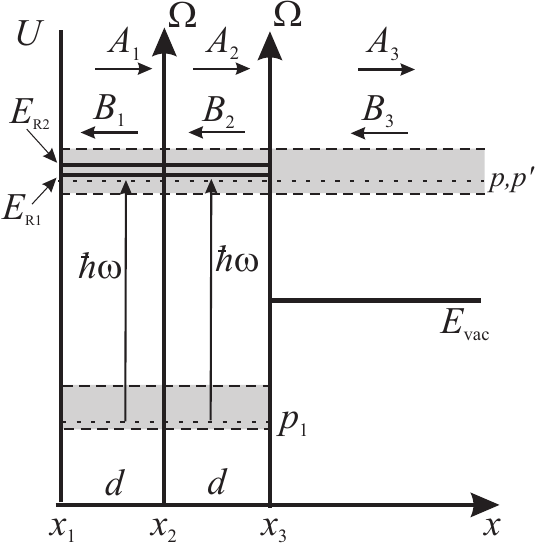}
\caption{Double-well heterostructure at an impermeable potential wall as a photoemitter}\label{fig:FIG1}
\end{figure}

As the main object of this article, we will consider a model that describes the simplest case when a double-well heterostructure serving as a photoemitter is deposited on a dielectric substrate. Let's direct the $x$ axis across the surfaces of the substrate and the heterostructure. Let the substrate have a wide energy gap above the vacuum level, which overlaps the narrow energy band recorded by the detector, covering one resonant doublet of quasi-stationary states of the double quantum well, so that in the main approximation the substrate is impenetrable for photoexcited electrons recorded by the detector. We will describe this by means of an infinitely high vertical wall on the left boundary  $x = {\kern 1pt} {x_1} = {\kern 1pt} {\kern 1pt} 0$ of the double quantum well, assuming that  ${\psi _p}(r) = {\psi _0}(E,x) = 0$ at   $x \le {\kern 1pt} {x_1} = {\kern 1pt} {\kern 1pt} 0$. To simplify the calculations, we replaced the lattice potential acting on electrons with the potential of a rectangular step with a height  ${E_{{\rm{vac}}}}$ at   $x{\kern 1pt} {\kern 1pt}  = {\kern 1pt} {\kern 1pt} {x_3}$. The position of the bottom of such a potential is determined by the electron affinity  $\chi $ in the photoemitter. We will model the potential barriers of the heterostructure with two delta functions  $U(x) = \left( {{\hbar ^2}/2m} \right)\sum\nolimits_{n = 2}^3 {\Omega \delta \left( {x - {x_n}} \right)}$ of equal power  $\Omega $ at a distance  $d$ from each other at  ${x_2} = {\kern 1pt} {\kern 1pt} {\kern 1pt} d,{\kern 1pt} {\kern 1pt} {\kern 1pt} {x_3} = {\kern 1pt} {\kern 1pt} 2d$ (Fig.\ref{fig:FIG1}). 

It is convenient to count the energy of electrons from the vacuum level. When photoexcited by light of frequency  $\omega $, electrons make transitions between states  ${p_1}$ localized at  $0{\kern 1pt}  \le x \le {\kern 1pt} {x_3}$ inside the photoemitter below the vacuum level, and the states  $p$ and  $p'$ excited above the vacuum level are delocalized along the entire semi-axis  $x > {\kern 1pt} 0$.  It is obvious that in this case, to describe excited states of photoemission, we must use only the inverse LEED method without an alternative. The detector of normal to the surface alternating current photoemission is located on the right and must be configured to register electrons with energies  ${\varepsilon _p}$ and  ${\varepsilon _{p'}}$ from a narrow band  ${E_{\min }} \le {\varepsilon _p},{\varepsilon _{p'}} \le {E_{\max }}$ of states $p$ and  $p'$  , covering one doublet of resonant quasi-stationary states with energies  ${E_{R1}}$ and  ${E_{R2}}$ above the vacuum level of the photoemitter. Electrons are effectively excited into the states of such a band if their initial states  ${p_1}$ belong to some also narrow energy band  ${E_{1\min }} \approx {E_{\min }} - \hbar \omega  \le {\varepsilon _{{p_1}}} \le {E_{1\max }} \approx {E_{\max }} - \hbar \omega $  below the vacuum level and the boundary level  ${E_0}$ (for a metal photoemitter this is the Fermi energy of a partially filled conduction band, and for a semiconductor photoemitter this is the energy of the top of the valence band)

For excited states of such a model, Fig.\ref{fig:FIG1} solutions  ${\psi _p}(r) = {\psi _j}(E,x)$   of the one-dimensional stationary Schrodinger equation at energy  $E > {E_{{\rm{vac}}}}$ in the regions $j = 1\,{\kern 1pt}$ (at  $0 < x < d$), $j = 2$ (at $d < x < 2d$),  $j = 3$ (at  $x > 2d$)  have the form 
\begin{equation}
\label{eq:math:eq5}
{\psi _j}(E,x) = {A_j}{e^{i{k_j}(x - {x_j})}} + {B_j}{e^{ - i{k_j}(x - {x_j})}}
\end{equation} 
where  ${k_j} = {\hbar ^{ - 1}}\sqrt {2m(E - {U_j})}$  is the wave number,   $m$ - electron mass,  ${U_j}$ -  potential energy of the electron in the  $j$-th region, $\;{k_1} = {k_2}$;  ${A_j}$ and   ${B_j}$ are partial amplitudes of plane monochromatic waves propagating to the right and left, respectively.

The transfer matrix method \cite{bib07},\cite{bib08},\cite{bib37}-\cite{bib38} allows us to relate the partial amplitudes of  three regions by linear relations

\begin{equation}
\label{eq:math:eq6}
\left( {\begin{array}{*{20}{c}}
{{A_2}}\\
{{B_2}}
\end{array}} \right) = L_2^{ - 1}{M_\Omega }M{L_1}\left( {\begin{array}{*{20}{c}}
{{A_1}}\\
{{B_1}}
\end{array}} \right),	
\end{equation}
\begin{equation}
\label{eq:math:eq7}
\left( {\begin{array}{*{20}{c}}
{{A_3}}\\
{{B_3}}
\end{array}} \right) = \tilde m\left( {\begin{array}{*{20}{c}}
{{A_1}}\\
{{B_1}}
\end{array}} \right),	
\end{equation}
\begin{equation}
\label{eq:math:eq8}
\tilde m = L_3^{ - 1}{M_\Omega }M{M_\Omega }M{L_1} = \left( {\begin{array}{*{20}{c}}
{{m_{11}}}&{{m_{12}}}\\
{{m_{21}}}&{{m_{22}}}
\end{array}} \right),
\end{equation}
\[{m_{11}} = m_{22}^ * ,{m_{12}} = m_{21}^ *\].

In accordance with the rigorous approach of the quantum theory of the atomic photoelectric effect \cite{bib23}-\cite{bib25} (starting with the pioneering work of M. Stobbe \cite{bib23}) and photoemission from crystals \cite{bib26}-\cite{bib31}, solutions of the Schrodinger equation, which describe the time-reversed boundary scattering problem, should be taken as the basis wave functions of photoexcited electrons. This means that in \eqref{eq:math:eq5} the partial amplitude  ${A_3}$ of the wave moving from the heterostructure to the right towards the detector must be specified by the normalization condition, which means it does not have pole features. Here it is convenient for us to specify it by the expression  ${A_3} = {\hbar ^{ - 1}}\sqrt {m/2\pi {k_3}} $, which ensures the normalization of the wave function  ${\psi _p}(r)$ to the delta function of energy \cite{bib01},\cite{bib36}, through which all other partial amplitudes can be expressed. In addition, in this case, the condition  ${B_1} =  - {A_1}$ must be satisfied to ensure that the wave function on the vertical wall is equal to zero  ${\psi _1}(E,0) = 0$. Then from \eqref{eq:math:eq7} we obtain
\begin{equation}
\label{eq:math:eq9}
{A_1} =  - {B_1} = \frac{{{A_3}}}{{{m_r}(E)}},
\end{equation}
\begin{equation}
\label{eq:math:eq10}
 {B_3} =  - \frac{{m_r^ * (E)}}{{{m_r}(E)}}{A_3} = \frac{1}{{r(E)}}{A_3}
 \end{equation}
where the designation is introduced
\begin{equation}
\label{eq:math:eq11}
 {m_r}(E) \equiv {m_r}(p) \equiv {m_{11}} - {m_{12}} = m_{22}^ *  - m_{21}^ * ,
  \end{equation}
and $r(E) = {{{A_3}} \mathord{\left/
 {\vphantom {{{A_3}} {{B_3}}}} \right. \kern-\nulldelimiterspace} {{B_3}}} =  - {{{m_r}(E)} \mathord{\left/
 {\vphantom {{{m_r}(E)} {m_r^ * (E)}}} \right.  \kern-\nulldelimiterspace} {m_r^ * (E)}}$  is the ''reflection'' amplitude,  
 $\left| r \right| = 1$, then from \eqref{eq:math:eq6} we express  ${A_2}$ and  ${B_2}$ through  ${A_3}$. All partial amplitudes (except ${A_3}$) can have pole singularities in the complex energy plane, which are determined by the denominators in \eqref{eq:math:eq9} and \eqref{eq:math:eq10} being zero  ${m_r}(E) = 0$. For a double quantum well, the complex roots  ${E_R} = {\mathop{\rm Re}\nolimits} {E_R} + i{\mathop{\rm Im}\nolimits} {E_R}$ of the equation  ${m_r}(E) \equiv {m_{11}} - {m_{12}} = m_{22}^ *  - m_{21}^ *  = 0$ are grouped into doublets. The real parts of the roots provide the energies of quasi-stationary levels, and their imaginary parts determine the spectral width and lifetimes of these quasi-stationary states. 

When calculating the resonant photocurrent of interest to us from a double quantum well using formulas \eqref{eq:math:eq1},  \eqref{eq:math:eq2}, it is necessary to sum over the excited states $p$ and  $p'$, the energies of which  $E$ and   $E'$ belong to the narrow band  ${E_{\min }} \le E,E' \le {E_{\max }}$ registered by the detector, covering a doublet of mutually close quasi-stationary levels  ${E_{R1}}$ and  ${E_{R2}}$ above the vacuum level, so that the boundaries  ${E_{\min }}$ and  ${E_{\max }}$ are sufficiently far from other quasi-stationary levels. 	Next, in the usual way, introducing multipliers equal to the energy densities of states  ${g_p} = dN/d{\varepsilon _p} = {\kern 1pt} {\kern 1pt} {\kern 1pt} dN/dE$, we replaced the summations over states  $p$ and  $p'$ to the numerical integration over the energies of these states  $E$ and   $E'$
\begin{equation}
\label{eq:math:eq12}
n(x,t) = 2\iint\limits_S {{\rho _{p',p}}(t){n_{p,p'}}(x){g_p}{g_{p'}}dEdE'},
\end{equation}
\begin{equation}
\label{eq:math:eq13}
j(x,t) = 2\iint\limits_S {{\rho _{p',p}}(t){j_{p,p'}}(x){g_p}{g_{p'}}dEdE'},  
\end{equation}
Here  ${n_{p,p'}}(x)$ and  ${j_{p,p'}}(x)$  are given \eqref{eq:math:eq3}, \eqref{eq:math:eq4}, integration is performed over the area   $S$  of the square   ${E_{\min }} \le E,E' \le {E_{\max }}$ on the   $E,E'$-plane. 

The density matrix satisfies the well-known kinetic equation \cite{bib11}-\cite{bib13}, which takes into account the influence of the external pumping electromagnetic field and the interaction of electrons with surrounding particles. This equation can be solved in different approximations. In \cite{bib02}, we wrote out such an equation and obtained formulas expressing its solutions in the dipole approximation in the second order of perturbation theory in the high-frequency electric field   ${E_\omega }$ of a light wave of frequency   $\omega $ after a sharp switching on of this field at the moment of time   $t = 0$ or a sharp switching off of the field at the moment of time   $t = {t_0}$. The effect of inelastic electron scattering inside the photoemitter is described in the usual way in the mass operator approximation (in the relaxation time approximation)  by means of effective complex additions to the electron energy, describing the electron-electron and electron-phonon interaction, which leads to a renormalization of energy levels, that is, to their shift   $\Delta {\varepsilon _p}$ and blurring  ${\gamma _p}$. These formulas \cite{bib02}  for  ${\rho _{p',p}}(t) \approx \rho _{p',p}^{(2)}(t)$, as usual, have the form of sums over the initially occupied unexcited states   ${p_1}$  of the electron inside the photoemitter effectively belonging also to a narrow energy band  ${E_{1\min }} \approx {E_{\min }} - \hbar \omega  \le {\varepsilon _{{p_1}}} \le {E_{1\max }} \approx {E_{\max }} - \hbar \omega $  in a partially filled quasi-two-dimensional conduction band or in a quasi-two-dimensional valence band of a photoemitter. The terms of the sums contain time exponents and frequency-energy denominators, as well as coefficients 
\begin{equation}
\label{eq:math:eq14}
{D_{{p_1}}} = \frac{{{n_{{p_1}}}}}{{{\hbar ^2}}}\left( {{d_{p',{p_1}}}{E_{ - \omega }}} \right)\left( {{d_{{p_1},p}}{E_\omega }} \right),	
\end{equation}
which are proportional to the light intensity  ${\left| {{E_\omega }} \right|^2}$, quasi-equilibrium occupation numbers of unexcited states  ${n_{{p_1}}}$  and the products of the components of matrix elements  ${d_{p',{p_1}}}$ and   ${d_{{p_1},p}}$ of  the electron dipole moment
\begin{equation}
\label{eq:math:eq15}
{d_{{p_1},p}}{\kern 1pt} {\kern 1pt}  = d_{p,{p_1}}^ *  = \int {\psi _{{p_1}}^ * ({\bf{r}}){\kern 1pt} {\kern 1pt} e{\bf{r}}{\psi _p}({\bf{r}})} {d^3}{\bf{r}}.      	 
\end{equation}
These parameters significantly determine the absolute value of the charge and current densities of photoelectrons.

In the model under study, Fig.\ref{fig:FIG1}, the matrix element   ${d_{{p_1},p}}$ of the electron dipole moment  \eqref{eq:math:eq15} necessarily has the above-mentioned pole features, since the wave functions of unexcited states  ${\psi _{{p_1}}}(r)$ are localized inside the photoemitter and effectively limit the integration in  \eqref{eq:math:eq15} to the region   ${x_1} = {\kern 1pt} {\kern 1pt} 0 < x < {x_3} = {\kern 1pt} {\kern 1pt} 2d$, in which the wave functions of excited states   ${\psi _p}(r)$ are characterized by partial coefficients   ${A_1}$,    ${B_1}$,   ${A_2}$ and   ${B_2}$. Taking out the pole factors from expressions  \eqref{eq:math:eq15}, we present   ${d_{{p_1},p}}$ and   ${d_{p',{p_1}}}$ in the form
\begin{equation}
\label{eq:math:eq16}
{d_{{p_1},p}} = \frac{{d_{{p_1},p}^0}}{{{m_r}(p)}},\quad      {d_{p',{p_1}}} = \frac{{d_{p',{p_1}}^0}}{{m_r^ * (p')}}, 	
\end{equation}
where   $d_{p',{p_1}}^0$ and  $d_{{p_1},p}^0$  are smooth functions of the energies of the excited electron. Consequently, according to  \eqref{eq:math:eq14}, all terms expressing   $\rho _{p',p}^{(2)}(t)$ sums over initially occupied states   ${p_1}$ are inversely proportional to the products   ${m_r}(p)m_r^ * (p')$, so it is convenient to represent the factor  ${D_{{p_1}}}$ in the form
\begin{equation}
\label{eq:math:eq17}
{D_{{p_1}}} = \frac{1}{{{m_r}(p)m_r^ * (p')}}D_{{p_1}}^0,
\end{equation}
where the multiplier
\begin{equation}
\label{eq:math:eq18}  
D_{{p_1}}^0 = \frac{{{n_{{p_1}}}}}{{{\hbar ^2}}}\left( {d_{p',{p_1}}^0{E_{ - \omega }}} \right)\left( {d_{{p_1},p}^0{E_\omega }} \right)
\end{equation}
is a smooth function of the energies of electrons in states   $p$ and   $p'$. 

The results of our numerical calculations show that with a sufficient width of the energy bands   $\left[ {{E_{\min }},{E_{\max }}} \right]$,   $\left[ {{E_{1\min }},{E_{1\max }}} \right]$, compared with the distance between the resonant levels $E_{R1}$  and $E_{R2}$  of the doublet of quasi-stationary states, the oscillating difference spatiotemporal component of the modulated photoemission pulse of interest to us over a wide range is qualitatively and quantitatively not very sensitive to the choice of boundaries   $\left[ {{E_{\min }},{E_{\max }}} \right]$ and  $\left[ {{E_{1\min }},{E_{1\max }}} \right]$. In these narrow integration bands, the parameters   ${E_\omega }$,  ${g_p},{g_{p'}},{g_{{p_1}}}$ and  $D_{{p_1}}^0 \approx D$ are almost constant factors. In addition, in [2] it was proven that under these conditions it is possible to analytically calculate with acceptable accuracy the main resonant contribution to the integrals that express sums over states   ${p_1}$ and obtain expressions for   $\rho _{p',p}^{(2)}(t)$ which are quite simple and convenient for further calculations:
for the pumping process at   $t \ge 0$
\begin{equation}
\label{eq:math:eq19}
\begin{array}{l}
\rho_{p',p}^{(2)}(t) = \cfrac{{2\pi i{\hbar ^2}D}}{{(({\kern 1pt} {\xi _{p'}} - {\xi _p}) + i{\gamma _{p'p}})}}\\
\quad  \times \cfrac{1}{{{m_r}(p)m_r^ * (p')}}{\kern 1pt} \left[ {1 - {e^{i({\xi _{p'}} - {\xi _p}){t \mathord{\left/
 {\vphantom {t \hbar }} \right.
 \kern-\nulldelimiterspace} \hbar } - {\gamma _{p'p}}{t \mathord{\left/
 {\vphantom {t \hbar }} \right.
 \kern-\nulldelimiterspace} \hbar }}}} \right]
\end{array}	
\end{equation}
where   ${\xi _p} - {\xi _{p'}} = E - E'$ and   ${\xi _p} - {\xi _{p'}} = E - E'$,
in the stationary saturation mode at   $t \to \infty $ this gives
\begin{equation}
\label{eq:math:eq20}
\rho _{p',p(0)}^{(2)} = \frac{{2\pi i{\hbar ^2}D}}{{(({\kern 1pt} {\xi _{p'}} - {\xi _p}) + i{\gamma _{p'p}})}}\frac{1}{{{m_r}(p)m_r^ * (p')}}, 
\end{equation}
and after a sharp switching off the pumping at the moment of time   ${t_0}$for   $t > {t_0}$
\begin{equation}
\label{eq:math:eq21}
\rho _{p',p}^{(2)}(t) = \rho _{p',p}^{(2)}({t_0}){e^{i({\xi _{p'}} - {\xi _p}){{(t - {t_0})} \mathord{\left/
 {\vphantom {{(t - {t_0})} \hbar }} \right.
 \kern-\nulldelimiterspace} \hbar } - {\gamma _{p'p}}{{(t - {t_0})} \mathord{\left/
 {\vphantom {{(t - {t_0})} \hbar }} \right.
 \kern-\nulldelimiterspace} \hbar }}},	
\end{equation}
where  $\rho _{p',p}^{(2)}({t_0})$ is the initial value arbitrarily specified at the moment   ${t_0}$, which can be estimated by expressions \eqref{eq:math:eq19} or \eqref{eq:math:eq20}. These expressions did not include the frequency  $\omega $  of light due to the rapid convergence of the integrals approximating the sums by states   ${p_1}$ for  $\rho _{p',p}^{(2)}(t)$. Here, in comparison with the corresponding formulas \cite{bib02}, a technical typo is corrected and the indices   $p$ and   $p'$ are rearranged.  Substitution \eqref{eq:math:eq19}-\eqref{eq:math:eq21} into \eqref{eq:math:eq1} and \eqref{eq:math:eq2} gives practically the same oscillatory-relaxation dependence of the charge and current densities of photoelectrons on time and coordinates as substitution of strict sums over states   ${p_1}$.

The general expressions for  $\rho _{p',p}^{(2)}(t)$,   $n(r,t)$, and   ${\bf{j}}(r,t)$ , derived in \cite{bib02} in the relaxation time approximation are applicable to a photoemitter in which the processes of inelastic electron scattering are weak, i.e. the thickness of the region of photoexcitation of electrons is less than the mean free path of high-energy electrons. Better than for a bulk photoemitter, our general formulas are applicable to photoemission from a separate double quantum well, which is a very thin photoemitter, the thickness of which is less than the mean free path of electrons.  
 Note also that the exact values of charge and current densities must be related by the continuity equation   $div{\bf{j}}({\bf{r}},t) =  - {{\partial n({\bf{r}},t)} \mathord{\left/
 {\vphantom {{\partial n({\bf{r}},t)} {\partial t}}} \right.
 \kern-\nulldelimiterspace} {\partial t}}$. However, the quantities   $n(x,t)$ and   $j(x,t)$ (the   $x$-th component of the current density vector) that we calculate, which are expressed by sums  \eqref{eq:math:eq1},  \eqref{eq:math:eq2} and integrals  \eqref{eq:math:eq12},  \eqref{eq:math:eq13}, satisfy the  one-dimensional equality
\begin{equation}
\label{eq:math:eq22} 
 \frac{{\partial j(x,t)}}{{\partial x}} \approx  - \frac{{\partial n(x,t)}}{{\partial t}} 	
\end{equation}
only approximately as long as the derivatives in \eqref{eq:math:eq22} are not small, both due to current leaks in the direction perpendicular to the  $x$-axis during inelastic scattering of electrons, and due to the approximations used for the density matrix (the initial representation of statistical averages as sums of products of time functions and coordinate functions and the calculation of the density matrix using perturbation theory). For the same reason, in the limit \eqref{eq:math:eq20} of the steady-state pumping regime with its infinite duration   ${t_0} = \infty $, calculations according to  \eqref{eq:math:eq12},  \eqref{eq:math:eq13} give the result   ${{\partial n(x,t)} \mathord{\left/
 {\vphantom {{\partial n(x,t)} {\partial t = 0}}} \right.
 \kern-\nulldelimiterspace} {\partial t = 0}}$, but    ${{\partial j(x,t)} \mathord{\left/
 {\vphantom {{\partial j(x,t)} {\partial x}}} \right.
 \kern-\nulldelimiterspace} {\partial x}} \ne 0$ at a finite distance   $x > 0$, so that   ${{\partial j(x,t)} \mathord{\left/
 {\vphantom {{\partial j(x,t)} {\partial x}}} \right.
 \kern-\nulldelimiterspace} {\partial x}} \to 0$ at   $x \to \infty $.

\section{Analytical methods for comparing the contributions of ongoing and incoming waves}\label{sec:sec3A}

In the inverse LEED photoemission picture we are considering, the formation of an oscillating electron wave packet occurs due to the abrupt turning on or off of electron photoexcitation inside the heterostructure. The width of the spatial localization of this wave packet is much larger than the width of the heterostructure, and its time evolution is determined by the dynamics of light pumping, by the oscillations of the electron wave function inside the heterostructure and the oscillations of the tunnel exit from this heterostructure, as well as by damping due to dissipation and by spreading due to dispersion.

The sections following this section present the results of numerical integration \eqref{eq:math:eq12}, \eqref{eq:math:eq13}, which confirm that charge and current density waves can form to the right of the heterostructure at  $x > 2d$, and in the studied scenario of inverse LEED photoemission, the contribution of the  ${B_3}$-waves coming from the detector is small compared to the contribution of the outgoing   ${A_3}$-waves. A rigorous analytical proof of these results is difficult, since the double integrals \eqref{eq:math:eq12}, \eqref{eq:math:eq13} cannot be represented as a product of two independent single integrals, as is done for pure quantum-mechanical states \cite{bib01},\cite{bib02}. We can only give some mathematical arguments about the relative magnitude and structure of the different contributions to these integrals in the simplest case of the validity of expressions \eqref{eq:math:eq19}-\eqref{eq:math:eq21} for the density matrix.

First of all, taking into account expressions \eqref{eq:math:eq5}, \eqref{eq:math:eq10}, we see from \eqref{eq:math:eq12} and \eqref{eq:math:eq13} that, to the right of the heterostructure at   $x > 2d$, the time-dependent contributions to the charge density   $\Delta n(x,t)$ and current density   $\Delta j(x,t)$ (determined by the time-dependent terms \eqref{eq:math:eq19}, \eqref{eq:math:eq21}) in the leading approximation are proportional to integral expressions of the form
\begin{equation}
\label{eq:math:eq23} 
\begin{array}{l}
\iint\limits_{S_k} {\left( {a_3^ + {e^{i({k_3}\tilde x - {{E\tilde t} \mathord{\left/
 {\vphantom {{E\tilde t} \hbar }} \right.
 \kern-\nulldelimiterspace} \hbar })}} + b_3^ + {e^{ - (i{k_3}\tilde x + {{E\tilde t} \mathord{\left/
 {\vphantom {{E\tilde t} \hbar }} \right.
 \kern-\nulldelimiterspace} \hbar })}}} \right)} \\
 \times \left( {a_3^ - {e^{ - i({{k'}_3}\tilde x - {{E'\tilde t} \mathord{\left/
 {\vphantom {{E'\tilde t} \hbar }} \right.
 \kern-\nulldelimiterspace} \hbar })}} + b_3^ - {e^{i{{k'}_3}\tilde x + {{E'\tilde t} \mathord{\left/
 {\vphantom {{E'\tilde t} \hbar }} \right.
 \kern-\nulldelimiterspace} \hbar }}}} \right)d{k_3}d{{k'}_3}
\end{array}			
\end{equation}
where   $\tilde t = t$ at   $0 < t < {t_0}$ and   $\tilde t = t - {t_0}$ at   $t > {t_0}$  (${t_0}$ is the duration of the pump pulse, which is almost rectangular in time) ;   $\tilde x = x - 2d$;  $E = {{{\hbar ^2}k_3^2} \mathord{\left/
 {\vphantom {{{\hbar ^2}k_3^2} {2m}}} \right.  \kern-\nulldelimiterspace} {2m}}$, 
 $E' = {{{\hbar ^2}{k'}_3^2} \mathord{\left/ {\vphantom {{{\hbar ^2}{k'}_3^2} {2m}}} \right.
 \kern-\nulldelimiterspace} {2m}}$;   
$a_3^ +  = a_3^ + (E,E') \propto {A_3}(E)$,   $b_3^ +  = b_3^ + (E,E') \propto {B_3}(E)$,   $a_3^ -  = a_3^ - (E,E') \propto A_3^ * (E')$,   $b_3^ -  = b_3^ - (E,E') \propto B_3^ * (E')$ are complex-valued functions of their arguments, including time-decay factors   ${e^{ - 2{\gamma _p}\tilde t/\hbar }}$, pole factor 
${({m_r}(E)m_r^ * (E'))^{ - 1}}$, and the energy denominator   $({\xi _{p'}} - {\xi _p}) + i{\gamma _{p'p}} = (E' - E) + i2{\gamma _p}$. The integration is performed over the area   ${S_k}$ of a square  ${\hbar ^{ - 1}}\sqrt {2m{E_{\min }}}  \equiv {k_{\min }} \le {k_3}$,   ${k'_3} \le {k_{\max }} \equiv {\hbar ^{ - 1}}\sqrt {2m{E_{\max }}}$ on the   ${k_3},{k'_3}$-plane.

For each of the two repeated integrals in \eqref{eq:math:eq23}, asymptotic estimates can be made using the fastest descent method as described in \cite{bib01}. In this case, in \eqref{eq:math:eq23} it is clearly seen that (without taking into account the pole factors) to the right of the heterostructure at positive values of   $\tilde x = x - 2d > 0$ for outgoing waves (the first terms in brackets) the points of the extreme phase   ${k_{3s}}{\kern 1pt}  = {k'_{3s}} = {{m\tilde x} \mathord{\left/
 {\vphantom {{m\tilde x} {\;\hbar \tilde t{\kern 1pt} {\kern 1pt}  > 0}}} \right.
 \kern-\nulldelimiterspace} {\;\hbar \tilde t{\kern 1pt} {\kern 1pt}  > 0}}$ of the exponential factor are located at positive values of the time parameter    $\tilde t > 0$, and for incoming waves (the second terms in brackets) the points of the extreme phase   ${k_{3s}}{\kern 1pt}  = {k'_{3s}} =  - {{m\tilde x} \mathord{\left/
 {\vphantom {{m\tilde x} {\;\hbar \tilde t{\kern 1pt} {\kern 1pt}  > 0}}} \right.
 \kern-\nulldelimiterspace} {\;\hbar \tilde t{\kern 1pt} {\kern 1pt}  > 0}}$ of the exponential factors are located at negative values of the time parameter   $\tilde t < 0$ . It follows that the resulting wave packet, which describes the photoemission current to the right of the heterostructure at   $\tilde x > 0$ and   $\tilde t > 0$, is mainly formed by waves outgoing to the right, and waves coming from the right can make a rather small contribution to it. This will be confirmed numerically below in 
Section 6. This picture is radically different from the picture of the evolution of a wave packet arriving from the right, which could be prepared at   $\tilde x > 0$ in some initial moment of time   $\tilde t < 0$ (with a predominant contribution of incoming waves), could reach the heterostructure at   $\tilde t \approx 0$, and then at   $\tilde t >0$ be transformed into a wave packet reflected from the heterostructure leaving to the right (with a predominant contribution of outgoing waves), a similar process is described in \cite{bib01} for the case of scattering of a wave packet incident on the heterostructure from the left. In all cases, when   $x$ and   $t$ change, the saddle points of the extreme phase and the associated fastest descent contours move, capturing nearby poles of   ${({m_r}(E)m_r^ * (E'))^{ - 1}}$, which can cause the appearance of wave-like oscillations of charge and current densities \cite{bib01},\cite{bib02}.

Using the saddle-point method as in \cite{bib02}, highlighting the main asymptotic contributions of the pole factors  ${({m_r}(E)m_r^ * (E'))^{ - 1}}$, it can be shown that in the region to the right of the double well at  $x > 2d$ these contributions  $\Delta n(x,t)$ and  $\Delta j(x,t)$ to both the charge and current densities have the character of weakly damped waves running to the right:
\begin{equation}
\label{eq:math:eq24} 
\begin{array}{l}
\Delta n(x,t) = \sum\limits_{j = 1}^2 {\left| {{{\tilde A}_{Rj}}} \right|{^2}} {e^{ - 2\left| {{{k''}_{Rj}}} \right|\tilde x - 2\left| {{{E''}_{Rj}}} \right|\tilde t}} + \,2\left| {{{\tilde A}_{R1}}{{\tilde A}_{R2}}} \right| \times \\
\quad \cos\left( {{\omega _{12}}\tilde t - {\kern 1pt} {\kern 1pt} {k_{12}}\tilde x + \,\Delta {\alpha _{12}}} \right){e^{ - (\left| {{{k''}_{R1}}} \right| + \left| {{{k''}_{R2}}} \right|)\tilde x - (\left| {{{E''}_{R1}}} \right| + \left| {{{E''}_{R2}}} \right|)\tilde t}}
\end{array}
\end{equation}
\begin{equation}
\label{eq:math:eq25} 
\begin{array}{l}
\Delta j(x,t) = \frac{\hbar }{m}{\kern 1pt} \left[ {\sum\limits_{j = 1}^2 {{{k'}_{Rj}}\left| {{{\tilde A}_{Rj}}} \right|{^2}} {e^{ - 2\left| {{{k''}_{Rj}}} \right|\tilde x - 2\left| {{{E''}_{Rj}}} \right|\tilde t}}} \right. + ({{k'}_{R1}} + {{k'}_{R2}}) \times \,\\
\left| {{{\tilde A}_{R1}}{{\tilde A}_{R2}}} \right|\left. {\cos\left( {{\omega _{12}}\tilde t  -   {k_{12}}\tilde x  + \Delta {\alpha _{12}}} \right){e^{ - (\left| {{{k''}_{R1}}} \right| + \left| {{{k''}_{R2}}} \right|)\tilde x - (\left| {{{E''}_{R1}}} \right| + \left| {{{E''}_{R2}}} \right|)\tilde t}}} \right],
\end{array}
\end{equation}
where  ${\omega _{12}} = 2\pi {\nu _{12}} = {\hbar ^{ - 1}}({E'_{R2}} - {E'_{R1}})$,  ${k_{12}} = {k'_{R2}} - {k'_{R1}}$ ,  $\Delta {\alpha _{12}} ={\alpha _{R1}} -  {\alpha _{R2}}$;  ${\tilde A_{Rj}} \equiv \tilde A({E_{Rj}})$ and  ${\alpha _{Rj}} = \alpha ({E_{Rj}})$ are amplitudes and phases of complex pre-exponential coefficients, determined by the residues of the integrand  \eqref{eq:math:eq23} at the poles   ${E_{Rj}} = {E'_{Rj}} + i{E''_{Rj}}$ ( $j = 1,2$)  of the corresponding partial coefficients; these poles correspond to complex wave numbers   ${k_{Rj}} \equiv k({E_{Rj}} ) = {\hbar ^{ - 1}}\sqrt {2m{E_{Rj}}}  = {k'_{Rj}} + i{k''_{Rj}}$ (we choose the branches of the roots so as to satisfy the physical conditions of wave attenuation in space). We are interested in systems that provide sufficiently slow attenuation, in which   ${E'_{Rj}} \gg \left| {{{E''}_{Rj}}} \right|$ and    ${k'_{Rj}} \gg \left| {{{k''}_{Rj}}} \right|$.

The integrand complex energy factor  ${\left( {(E' - E) + i2{\gamma _p}} \right)^{ - 1}} = {\left( {\sqrt {{{(E' - E)}^2} + {{(2{\gamma _p})}^2}} } \right)^{ - 1}}{e^{iarctg\left( {{{2{\gamma _p}} \mathord{\left/
 {\vphantom {{2{\gamma _p}} {(E - E')}}} \right.
 \kern-\nulldelimiterspace} {(E - E')}}} \right)}}$ also plays an important role, its modulus   $ \sim \gamma _p^{ - 1}$ is effectively large in a narrow band of  width   $ \sim {\gamma _p}$ and area   $ \sim {\gamma _p}({E_{\max }} - {E_{\min }})$ along the diagonal of the square of integration   ${E_{\min }} \le E,E' \le {E_{\max }}$, providing to the right of the heterostructure at  $x > 2d$ a large contribution of the quasi-diagonal terms with   $E \approx E'$ to the asymptotically steady-state current at   ${t_0} \to \infty $ (expression  \eqref{eq:math:eq20}) and determining at   $t \le {t_0}$ the values of the time-independent contributions to charge   $\Delta n(x)$ and current   $\Delta j(x)$ densities in the pump regime, related to the unit in square brackets  \eqref{eq:math:eq19}, and hence the maximum values of   $n(x,t)$ and  $j(x,t)$ at    $t \approx {t_0}$. Even more importantly, due to its complex phase, this factor additionally violates the phase symmetry between the outgoing and incoming waves in such a way that the contribution to the photoemission current of the   ${A_3}$-waves outgoing to the detector is greater than the contribution of the   ${B_3}$-waves incoming from the detector, although the moduli of their amplitudes, in accordance with  \eqref{eq:math:eq10}, are the same   $\left| {{B_3}} \right| = \left| {{A_3}} \right|$ due to the presence of an impenetrable vertical wall on the left. The reason for this can be understood by considering the integrals
\begin{equation}
\label{eq:math:eq26}
\begin{array}{l}
\iint\limits_{k,k'} {\cfrac{{f({k_3},{k'_3})}}{{k_3^2 -{ k'_3}^2 - i\alpha }}} {e^{i[(E' - E)\tilde t/\hbar \, \pm \,({k_3} - {k'_3})\tilde x]}}d{k_3}d{k'_3}\\
 = \iint\limits_{\kappa, \kappa'} {\cfrac{{\tilde f(\kappa ,\kappa ')}}{{\kappa \kappa ' - i\alpha }}} {e^{ - i(\hbar \kappa \kappa '\tilde t/2m\, \mp \,\kappa \tilde x)}}d\kappa d\kappa ',
\end{array} 
\end{equation}
where   $\alpha  = 4m{\gamma _p}/{\hbar ^2}$. Such integrals arise for the region  $x > 2d$ after substituting expressions \eqref{eq:math:eq5} and \eqref{eq:math:eq19}, \eqref{eq:math:eq21} or \eqref{eq:math:eq20} (when here   $\tilde t = 0$) into \eqref{eq:math:eq12} and \eqref{eq:math:eq13} from terms proportional to   ${A_3}A_3^ * $ (giving the contributions   $\Delta {n_{{B_3} = 0}}$,   $\Delta {j_{{A_3} = 0}}$ of only outgoing waves - the case of the plus sign in the exponent) and for terms proportional to   ${B_3}B_3^ * $ (giving the contributions   $\Delta {n_{{A_3} = 0}}$,  $\Delta {j_{{A_3} = 0}}$ of only incoming waves - the case of the minus sign in the exponent) and subsequent replacement of integration variables   $\kappa  = {k_3} - {k'_3}$,   $\kappa ' = {k_3} + {k'_3} > 0$ (in the left part of \eqref{eq:math:eq26} the integration is performed over a square   ${\hbar ^{ - 1}}\sqrt {2m{E_{\min }}}  \equiv {k_{\min }} \le {k_3}$,   ${k'_3} \le {k_{\max }} \equiv {\hbar ^{ - 1}}\sqrt {2m{E_{\max }}}$ on the   ${k_3},{k'_3}$-plane), and in the right part of \eqref{eq:math:eq26} - over a square rotated by   ${45^ \circ }$ with vertices at points   $(0,2{k_{\min }})$,   $({k_{\max }} - {k_{\min }},{k_{\min }} + {k_{\max }})$,   $(0,2{k_{\max }})$,   $({k_{\min }} - {k_{\max }},{k_{\min }} + {k_{\max }})$ on the   $\kappa ,\kappa '$-plane). Functions   $f({k_3},{k'_3})$ and   $\tilde f(\kappa ,\kappa ')$ include all other factors. By equating to zero the derivative with respect to   $\kappa $ of the phase   $\varphi  = {\mathop{\rm arctg}\nolimits} (\alpha /\kappa \kappa ') - \hbar \kappa \kappa '\tilde t/2m\, \pm \,\kappa \tilde x$ of the integrand \eqref{eq:math:eq26} (neglecting the influence of the phase  $\tilde f(\kappa ,\kappa ')$), we obtain the condition for the extreme  phase    $\tilde x =  \pm (\hbar \kappa '/2m)(\tilde t + 2m\alpha /({\alpha ^2} + {(\kappa \kappa ')^2})$. For   $\tilde t >;0$ (or   $\tilde t = 0$) and   $\tilde x > 0$ it can be satisfied only with the plus sign, in which case, for a fixed   $\kappa ' > 0$, the integral over   $\kappa $ must be significantly larger than it would be with the minus sign.

The contributions of the time-independent terms \eqref{eq:math:eq19}, \eqref{eq:math:eq20} are given by \eqref{eq:math:eq26}, in which we should put   $\tilde t =  0$, in which case we can confirm the previous statement by noting that for each fixed value of    $\kappa ' > 2{k_{\min }} > 0$ the integrand of the integral over   $\kappa $  has a pole  ${\kappa _0} = {{i4m{\gamma _p}} \mathord{\left/
 {\vphantom {{i4m{\gamma _p}} {\kappa '{\hbar ^2}}}} \right.
 \kern-\nulldelimiterspace} {\kappa '{\hbar ^2}}}$ in the upper half-plane of the complex variable . The contribution of this pole is large in the case of the plus sign (when the auxiliary integration contour on the complex   $\kappa $-plane must be formally expanded and closed in the upper half-plane) and small in the case of the minus sign (when the integration contour must be closed in the lower half-plane). This is obvious if we assume that in the vicinity of the extreme phase point  and the pole $\kappa_0$  the influence of factors  ${({m_r}(E)m_r^ * (E'))^{ - 1}}$ is weak, so that the functions  $f({k_3},{k'_3})$ and  $\tilde f(\kappa ,\kappa ')$ are sufficiently smooth and analytical in this region.
 
The space-time dependence of the relatively small contributions $\Delta {n_{{A_3} = 0}}$,  $\Delta {j_{{A_3} = 0}}$ of incoming  ${B_3}$-waves can be estimated by taking into account that the corresponding exponentials in \eqref{eq:math:eq23} quickly oscillate with an average zero value at large values of  $x$ and  $\tilde t$. Their small contribution is determined by a narrow region of width  $\Delta E \ll ({E_{\max }} - {E_{\min }})$, so that integrals \eqref{eq:math:eq12} and \eqref{eq:math:eq13} become proportional to the product  $ \sim {\gamma _p}\Delta E \ll {\gamma _p}({E_{\max }} - {E_{\min }})$.  The energy width  $\Delta E = {{{\hbar ^2}(k_{3(2)}^2 - k_{3(1)}^2)} \mathord{\left/ {\vphantom {{{\hbar ^2}(k_{3(2)}^2 - k_{3(1)}^2)} {2m}}} \right.
 \kern-\nulldelimiterspace} {2m}}$ (where  ${k_{3(1)}}$ and   ${k_{3(2)}}$ are the values of the wave number  ${k_3}$ at the boundaries of this region) corresponds to a small part of the interval  ${E_{\max }} - {E_{\min }}$, on this part the phase of the exponential changes by a value of the order of   $\pi  \approx ({k_{3(2)}} - {k_{3(1)}})x + \Delta E\,\tilde t/\hbar $, from which we have  ${k_{3(2)}} - {k_{3(1)}} \approx \pi /(x + \tilde t\,\hbar k/m)$, where  $k = ({k_{3(2)}} + {k_{3(1)}})/2$,     ${k_{\min }} < k < {k_{\max }}$,  and therefore
\begin{equation}
\label{eq:math:eq27}
\Delta E  \approx  \frac{{\pi \hbar }}{{\tilde t + {{mx} \mathord{\left/
 {\vphantom {{mx} {\hbar k}}} \right.
 \kern-\nulldelimiterspace} {\hbar k}}}}  \approx  \left\{ {\begin{array}{*{20}{c}}
{\quad {{\pi \hbar } \mathord{\left/
 {\vphantom {{\pi \hbar } {\tilde t}}} \right.
 \kern-\nulldelimiterspace} {\tilde t}},}&{\tilde t \gg {{mx} \mathord{\left/
 {\vphantom {{mx} {\hbar k}}} \right.
 \kern-\nulldelimiterspace} {\hbar k}},(a)}\\
{{{({{\pi k\hbar } \mathord{\left/
 {\vphantom {{\pi k\hbar } m}} \right.
 \kern-\nulldelimiterspace} m})} \mathord{\left/
 {\vphantom {{({{\pi k\hbar } \mathord{\left/
 {\vphantom {{\pi k\hbar } m}} \right.
 \kern-\nulldelimiterspace} m})} {x}}} \right.
 \kern-\nulldelimiterspace} {x}},}&{\tilde t \ll {{mx} \mathord{\left/
 {\vphantom {{mx} {\hbar k}}} \right.
 \kern-\nulldelimiterspace} {\hbar k}}.(b)}
\end{array}} \right.
\end{equation}
Small contributions $\Delta {n_{{A_3} = 0}}$ and $\Delta {j_{{A_3} = 0}}$ are proportional to this value. The estimate described by line \eqref{eq:math:eq27} is also valid for $\tilde t = 0$, that is, for time-independent (in particular, for stationary, asymptotic for  ${t_0} = \infty $) contributions to  $\Delta {n_{{A_3} = 0}}$,  $\Delta {j_{{A_3} = 0}}$.
 
 The arguments and estimates presented in this section are not strict proofs; they are of an approximate indicative nature and are more suitable for cases of integration over a fairly wide energy band. In our case, the observed photoemission pulse and integration band are spectrally narrow. For the effects under discussion to manifest, it is sufficient that the poles and points of extreme phases responsible for them are located even outside this band, but close to it. As a result, the actual ratio of the discussed contributions of   ${A_3}$- and  ${B_3}$-waves to the charge and current densities may not be very large. But the results of numerical calculations given below generally confirm the indicated tendencies.

\section{Main results of model numerical calculations}\label{sec:sec4A}

\begin{figure}[h]
\vspace{-2ex}
\centering
\includegraphics[width=7 cm]{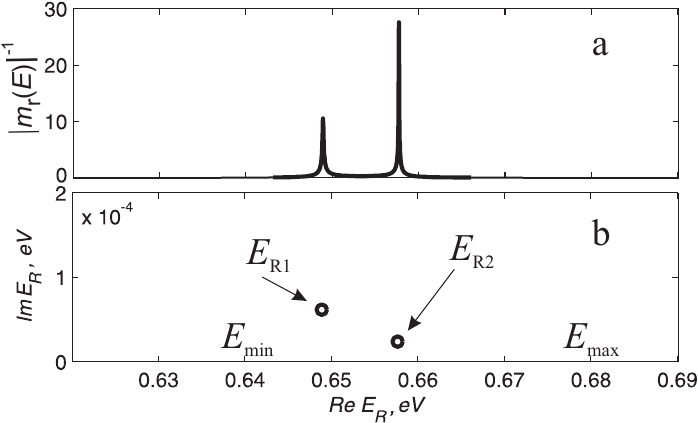}
\caption{For the studied doublet at   $\Omega  = 10$a.u. : a) the energy dependence of the value  ${\left| {{m_r}(E)} \right|^{ - 1}}$ and b) the poles of all partial amplitudes except  ${A_3}$ (zeros of the amplitude  $r(E)$ of  ''eflection'' from the heterostructure)}\label{fig:FIG2}
\end{figure}

Below we present the results of numerical simulation for a photoemitter with a surface heterostructure, a simplified energy scheme of which is shown in Fig.\ref{fig:FIG2}, with the following given parameters:  $d = 125$ \AA,  $\Omega  = 10$ a.u.  
$ = 18.9$ \AA$^{-1}$, $\chi  = 4$ eV. 

		By solving equation  ${m_r}(E) \equiv {m_{11}} - {m_{12}} = m_{22}^ *  - m_{21}^ *  = 0$ numerically, we found that the lower doublets above the vacuum level are located near the energies (0.036 eV, 0.044 eV), (0.236 eV, 0.244 eV), (0.440 eV, 0.448 eV), (0.649 eV, 0.658 eV), (0.863 eV, 0.871 eV),... . Difference oscillations of the photoemission charge and current densities can manifest a "wave packet" formed by the superposition of photoelectrons with energies from a certain band  ${E_{\min }} \le E \le {E_{\max }}$ wide enough to cover one doublet of resonant quasi-stationary states, but narrow compared to the distances to neighboring doublets. Such a pulse can be created by separating photoelectrons with energies  ${E_{\min }} \le E \le {E_{\max }}$ by means of magnetic and electric fields of the appropriate configuration.

By numerically integrating \eqref{eq:math:eq12} and \eqref{eq:math:eq13}, we calculated the photoemission charge and current densities formed by excited electrons whose energies belong to the energy band from  ${E_{\min }} = 0.640$ eV to  ${E_{\max }} = 0.680$ eV, covering the fourth above the vacuum level resonant doublet of quasi-stationary states of the double-well heterostructure, which corresponds to two close poles of all except  ${A_3}$ partial amplitudes (zeros of the ''reflection'' amplitude  $r(E)$), i.e. two roots of the equation  ${m_r}(E) \equiv {m_{11}} - {m_{12}} = m_{22}^ *  - m_{21}^ *  = 0$:  ${E_{R1}} = (0.6489 + i6.186 \cdot {10^{ - 5}})$eV and  ${E_{R2}} = (0.6576 + i2.390 \cdot {10^{ - 5}})$ eV. For this band, Fig.\ref{fig:FIG2}a shows the spectrum of the quantity  ${\left| {{m_r}(E)} \right|^{ - 1}}$, and Fig.\ref{fig:FIG2}b shows the position of the doublet of poles on the complex energy plane.

The heterostructure is practically impenetrable to electrons outside resonances, and the narrow resonance peaks of all partial amplitudes except  ${A_3}$  have a width of the order of the imaginary part of the poles. For the lifetimes of quasi-stationary states associated with this doublet, we have values  ${\tau _{R1}} \approx \hbar /|{E''_{R1}}| \approx 1.07 \cdot {10^{ - 11}}$s $ = 4.40 \cdot {10^5}$a.u.,  ${\tau _{R2}} \approx \hbar /|{E''_{R2}}|$ $ = 2.76 \cdot {10^{ - 11}}$s $ = 1.14 \cdot {10^6}$a.u., that is  ${\tau _{R1}} \approx {\tau _{R2}}$. The difference in energies of these states  $\Delta {E_{R12}} = {E'_{R2}} - {E'_{R1}} = 8.77 \cdot {10^{ - 3}}$eV determines the frequency  ${\nu _{12}} = \Delta {E_{R12}}/2\pi \hbar  \approx 2.10 \cdot {10^{12}}$Hz and the period  ${T_{12}} = 1/{\nu _{12}} = 4.73 \cdot {10^{ - 13}}$s $ \approx 1.95 \cdot {10^4}$a.u. of the photocurrent oscillations.

Current oscillations should be effectively observed when the inequalities  ${\tau _{R1}},{\tau _{R2}},{\tau _p} \gg {T_{12}}$ are satisfied, where  ${\tau _p} = \hbar /{\gamma _p}$ is the electron relaxation time determined by inelastic scattering. In the described numerical calculations we took the value  ${\gamma _p} = 1 \cdot {10^{ - 4}}$eV =  $3.675 \cdot {10^{ - 6}}$ a.u., i.e.  ${\tau _p} = 6.58 \cdot 10{}^{ - 12}$s  $ = 2.72 \cdot {10^5}$ a.u. This value is somewhat overestimated even for bulk semiconductors, and even more so for heterostructure thin-film conducting layers. We took it almost comparable with the imaginary parts of the poles  $|{\mathop{\rm Im}\nolimits} {E_{R1,2}}|$, which allows us to demonstrate on the calculated graphs of the spatial and temporal dependences of the charge and current densities in a visual scale both the damped resonant oscillations and the exit to the asymptotics of the stationary and zero current at large times. For a relatively long existence of difference oscillations it is advantageous for  ${\gamma _p}$ to be as small as possible; this can be expected in very thin quantum-dimensional films in which inelastic scattering of electrons is weaker than in the bulk.

By changing the boundary values   ${E_{\min }}$ and  ${E_{\max }}$ we have become convinced that for a sufficient width of the energy bands  $\left[ {{E_{\min }},{E_{\max }}} \right]$,  $\left[ {{E_{1\min }},{E_{1\max }}} \right]$ in comparison with the distance between the resonance levels  ${E_{R1}}$ and  ${E_{R2}}$ of the doublet of quasi-stationary states, the pole contribution of these states and the frequency-difference oscillatory space-time component of the modulated photoemission pulse of interest to us are qualitatively and quantitatively almost independent of the choice of boundaries  $\left[ {{E_{\min }},{E_{\max }}} \right]$ in a sufficiently wide range between neighboring doublets.     The absolute values of the photoemission charge and current densities can vary within very wide limits depending on the frequency and intensity  ${\left| {{E_\omega }} \right|^2}$ of light, as well as on the physicochemical nature of the photoemitter material and the structure of potential barriers. In narrow integration bands  ${E_{\min }} \le E,E' \le {E_{\max }}$, the parameters  ${E_\omega }$, ${g_p},{g_{p'}},{g_{{p_1}}}$, ${\gamma _p}$ and  $D_{{p_1}}^0 \approx D$ can be considered almost constant (if necessary, they can be estimated numerically).   We calculated the ratios of the photoemission current $j(x,t)$ and charge $n(x,t)$ densities (taking into account the relationship \eqref{eq:math:eq24}) to the number equal to the value of the current density  $j({x_3},{t_0})$ at the output $x = {x_3} = 250$ \AA  of the heterostructure with  $\Omega  = 10$ a.u. at the moment of turning off the pumping  ${t_0} = {10^5}$ a.u. =  $2.419 \cdot {10^{ - 12}}$ s. In such relations the dependence on the quantities ${E_\omega }$, ${g_p},{g_{p'}}$ and  $D$ almost cancels out, therefore in our calculations it was possible to consider these quantities formally equal to unity, at  $\Omega  = 10$ a.u. this gave the absolute values  $j({x_3},{t_0}) = 3.31 \cdot {10^{ - 3}}$ a.u. (unit value on the ordinate axes of all the corresponding graphs of this article) and  $n({x_3},{t_0}) = 1.52 \cdot {10^{ - 2}}$ a.u.

\begin{figure}[h]
\vspace{-2ex}
\centering
\includegraphics[width=7 cm]{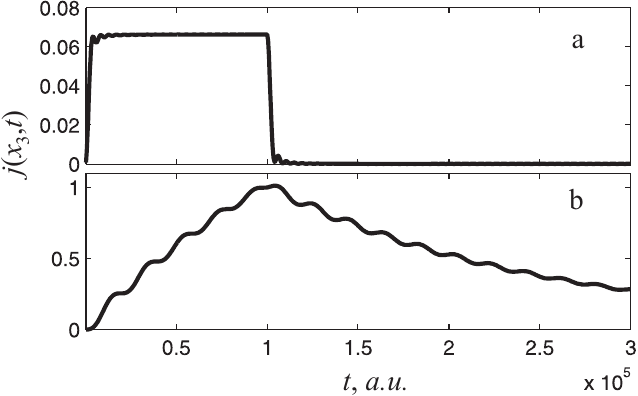}
\caption{Time dependence of the current density on the surface  $x = {x_3} = 2d$ of the photoemitter a) at $\Omega  = 0$, b) at  $\Omega  = 10$ a.u. and the duration of pumping  ${t_0} = {10^5}$ a.u.}\label{fig:FIG3}
\end{figure}

\begin{figure}[h]
\vspace{-2ex}
\centering
\includegraphics[width=7 cm]{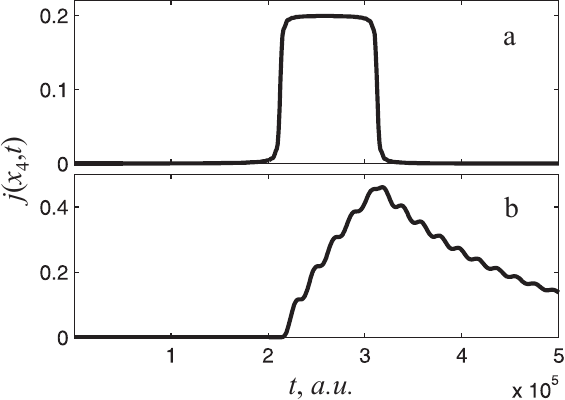}
\caption{Time dependence of the current density at the point $x = {x_4} = 2.5 \cdot {10^4}$ \AA:   a) at $\Omega  = 0$, b) at  $\Omega  = 10$ a.u. and the duration of pumping  ${t_0} = {10^5}$ a.u.}\label{fig:FIG4}
\end{figure}

\begin{figure}[h]
\vspace{-2ex}
\centering
\includegraphics[width=7 cm]{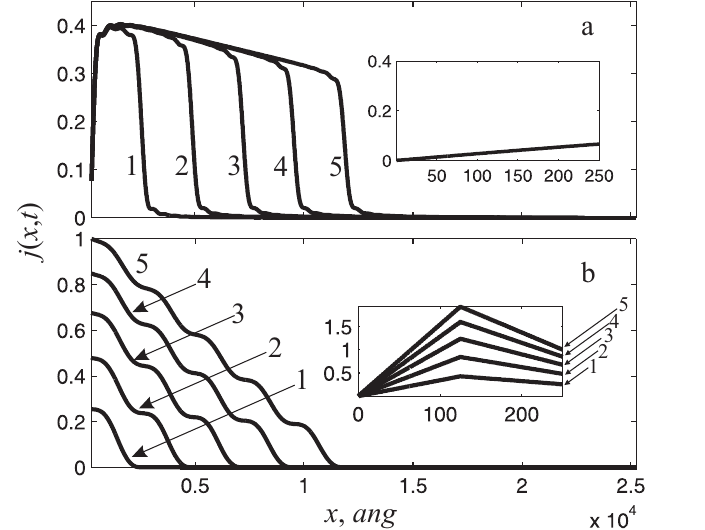}
\caption{a) for  $\Omega  = 0$ and b) for  $\Omega  = 10$ a.u. coordinate dependence of the current density at  $x > {x_3} = 2d$ and  $t \le {t_0} = {10^5}$ a.u. at the moments of time: 1)  $t = 2 \cdot {10^4}$ a.u., 2)  $t = 4 \cdot {10^4}$ a.u., 3)  $t = 6 \cdot {10^4}$ a.u., 4)  $t = 8 \cdot {10^4}$ a.u., 5)  $t = {t_0} = {10^5}$ a.u. The insets show this dependence inside the heterostructure at  $x < {x_3} = 2d$}\label{fig:FIG5}
\end{figure}

\begin{figure}[h]
\vspace{-2ex}
\centering
\includegraphics[width=6 cm]{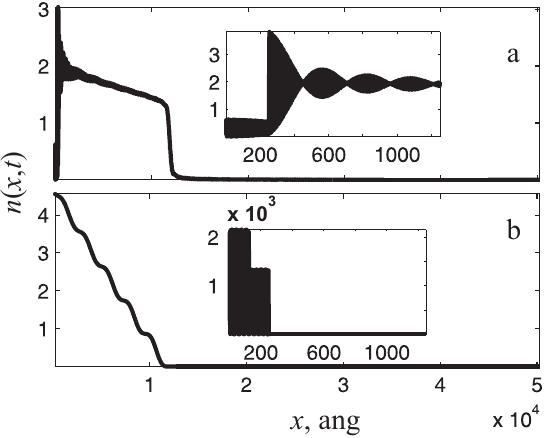}
\caption{a) for  $\Omega  = 0$ and b) for  $\Omega  = 10$a.u. the coordinate dependence of the charge density corresponding to the fifth pulse of the figures in Fig.\ref{fig:FIG5} at the moment  $t = {t_0} = {10^5}$ a.u. of the end of light pumping outside the heterostructure at $x > {x_3} = 2d$. The insets show the same on an extended scale at  $x < {x_3} = 2d$ inside and at $x > {x_3} = 2d$ near the heterostructure.}\label{fig:FIG6}
\end{figure}

\begin{figure}[h]
\vspace{-2ex}
\centering
\includegraphics[width=7 cm]{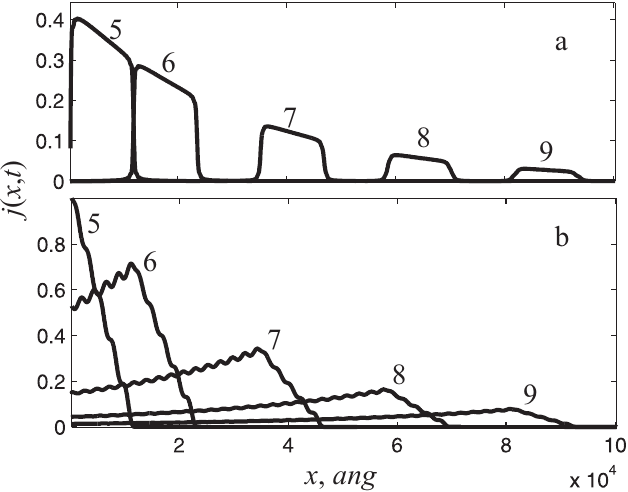}
\caption{ a) for  $\Omega  = 0$ and b) for  $\Omega  = 10$ a.u. coordinate dependence of the current density at  $x > {x_3} = 2d$ and  $t \ge {t_0} = {10^5}$ a.u. at the moments of time: 5)  $t = {t_0}\; = {10^5}$ a.u., 6)  $t = 2 \cdot {10^5}$ a.u., 7)  $t = 4 \cdot {10^5}$ a.u., 8)  $t = 6 \cdot {10^5}$ a.u., 9)  $t = 8 \cdot {10^5}$ a.u.}\label{fig:FIG7}
\end{figure}

Below in the figures \ref{fig:FIG3} - \ref{fig:FIG5} and Fig.\ref{fig:FIG7} of this section are given the results of calculations of the space-time dependences of the photoelectron current densities  $j(x,t)$ at to the right of the heterostructure  $x > {x_3} =  2d$, obtained by direct numerical integration of expressions \eqref{eq:math:eq12} and \eqref{eq:math:eq13} taking into account all four exponential factors \eqref{eq:math:eq23}, i.e.  ${A_3} \ne 0,{B_3} \ne 0$  for the pumping time  ${t_0} = {10^5}$ a.u. For all these cases except Fig.\ref{fig:FIG6}, a, the calculated curves for similar charge density dependencies  $n(x,t)$ are not given here, since on this scale they look the same as the curves  $j(x,t)$, differing from them in the main order by a practically constant factor   $n(x,t) = {\rm{v}}_g^{ - 1}j(x,t) = 4.566\,j(x,t)$, where  ${{\rm{v}}_g} = j({x_3},{t_0})/n({x_3},{t_0}) \approx 4.8 \cdot {10^5}$ m/s  $ \approx 0.219$ a.u.,  so that   $j(x,t) \approx {{\rm{v}}_g}n(x,t)$. The factor  ${{\rm{v}}_g} \equiv {{\hbar {k_g}} \mathord{\left/
 {\vphantom {{\hbar {k_g}} m}} \right.
 \kern-\nulldelimiterspace} m} \approx {{\hbar {k_3}} \mathord{\left/
 {\vphantom {{\hbar {k_3}} m}} \right.
 \kern-\nulldelimiterspace} m}$ before  $n(x,t)$ arises in the calculation $j(x,t)$ as a result of the action of the operator $ \nabla$  in the integral \eqref{eq:math:eq13} and is equal to the effective average velocity of photoelectrons in a narrow integration band; this velocity approximately corresponds to the energy  ${E_c} = m{\rm{v}}_g^2/2 = 0.653$ eV.

Figure \ref{fig:FIG3} shows the time dependence of the photoelectron current density  $j(x,t) = j({x_3},t)$ on the photoemitter surface $x = {x_3} = 2d$; a) in the absence of delta barriers  at  $\Omega  = 0$ (near rectangular pulse of duration $\approx t_0$) and b) at  $\Omega  = 10$ a.u. (the magnitude of which  $j({x_3},{t_0})$ is taken as the relative unit of current density).
It is noteworthy (Fig.\ref{fig:FIG3},a) that the magnitude of the current density is comparatively small  $j({x_3},t) \approx 0.06$ at  $\Omega  = 0$ in the absence of delta barriers due to the strong influence of the boundary condition  ${\psi _p}(0) = {\psi _1}(E,0) = 0$ on the impermeable wall at  $x = 0$ (Section 7  below). At a large distance from the photoemitter   $x = {x_4} = 2.5 \cdot {10^4}$\AA $ \gg {x_3} = 2d$, the time dependence of the current density looks like that shown in (Fig.\ref{fig:FIG4}): a) at  $\Omega  = 0$ in the absence of delta barriers, the current density is an order of magnitude greater $j({x_4},t) \approx 0.2$, and b) at  $\Omega  = 10$ a.u. it is approximately two times less  $j({x_4},t) \approx 0.5$ than at  $x = {x_3} = 2d$.

An important result is demonstrated by Fig.\ref{fig:FIG3},b and Fig.\ref{fig:FIG4},b, which show difference oscillations with a time period of  $T \approx 4.73 \cdot {10^{ - 13}}$s $ = 1.95 \cdot {10^4}$ a.u., what is close to  ${T_{12}} = 1/{\nu _{12}} = 4.76 \cdot {10^{ - 13}}$ s $ \approx 1.97 \cdot {10^4}$ a.u., the exponential decay time of these oscillations is of the order of  $\tau  = {\hbar  \mathord{\left/
 {\vphantom {\hbar  {\left( {\left| {{{E''}_{R1}}} \right| + \left| {{{E''}_{R2}}} \right|} \right)}}} \right.
 \kern-\nulldelimiterspace} {\left( {\left| {{{E''}_{R1}}} \right| + \left| {{{E''}_{R2}}} \right|} \right)}} = 7.68 \cdot {10^{ - 12}}$s $ = 3.17 \cdot {10^5}$ a.u., and the decay time of the entire current pulse is close to  ${\tau _p}$.

Figures \ref{fig:FIG5} illustrate the coordinate dependence of the increasing current density pulse  $j(x,t)$ for different moments of time during the process of light pumping  $t \le {t_0}$ outside the heterostructure at  $x > {x_3} = 2d$ (the insets show the spatial dependence $j(x,t)$ inside the heterostructure at  $x < {x_3} = 2d$): a) in the absence of delta barriers at  $\Omega  = 0$ and b) at  $\Omega  = 10$ a.u.. Figures \ref{fig:FIG6} show the corresponding coordinate dependence of the charge density $n(x,t)$ for the fifth pulse of the figures in Fig.\ref{fig:FIG5} at the moment  $t = {t_0} = {10^5}$ a.u. of the end of light pumping outside the heterostructure, and the insets show the same on an extended scale inside and near the heterostructure a) at  $\Omega  = 0$  and b) at  $\Omega  = 10$ a.u. 

Figure \ref{fig:FIG7} shows the coordinate dependence $j(x,t)$ calculated with the density matrix \eqref{eq:math:eq21}, after the abrupt switching off of the pump at  $t \ge {t_0}$ and $x > {x_3} = 2d$ a) at  $\Omega  = 0$ and b) at  $\Omega  = 10$a.u. It describes the propagation and strong relaxation damping of the electron pulse that is breaking away or has broken away from the photoemitter.

It is evident from the figures that, under the same light pumping conditions, the shape and magnitude of the photoemission pulse from a thin-film heterostructure with delta barriers (panels (b) at  $\Omega  \ne 0$) are very different from a similar pulse from a film of the same thickness without delta barriers (panels (a) at $\Omega  = 0$). At  $\Omega  \ne 0$ the magnitude of the photoemission pulse is greater than at $\Omega  = 0$, it is obvious that this increase in the pulse at  $\Omega  \ne 0$ is the result of the effective spatial restriction of the electron motion between the delta barriers inside the emitting heterostructure (where the electron concentration is three orders of magnitude greater than outside, as demonstrated by the inset in Fig.\ref{fig:FIG6},b). Note also that in the case of photoemission from a bulk photoemitter with a similar heterostructure on its surface \cite{bib02}, on the contrary, at  $\Omega  \ne 0$ the pulse magnitude is smaller than at  $\Omega  = 0$ as a result of the low transparency of the delta barriers.

During pumping  $t < {t_0}$ at  $\Omega  = 0$ (Fig. 5,b) the current density maxima are located on the left delta barrier inside the heterostructure and outside it on the heterostructure surface at  $x = {x_3} =  2d$, increasing with time. However, at   $\Omega  = 0$ (Fig.\ref{fig:FIG5},a) the current density smoothly increases from left to right inside the heterostructure with an almost identical slope at any  $t \le {t_0}$ has an almost constant value of  $j({x_3},t) \approx 0.06$ a.u. on its surface, oscillating near the heterostructure and reaching a maximum of  $j(x,t) \approx 0.4$ a.u. at  $x \approx {10^3}$ \AA, which is the result of interference of the outgoing and incoming partial waves. These interference oscillations are especially noticeable in the curves (Fig.\ref{fig:FIG6},a) of the coordinate dependence of the charge density  $n(x,t)$ during pumping: inside the film, oscillations of a nearly standing wave with a wavelength of several angstroms occur, and outside, near the surface, with a carrier wave length of  ${\lambda _g} = 2\pi /{k_g} \approx 15.27$ \AA (not resolved in the scale of the figure), but with an envelope in the form of a standing wave having a wavelength of  $ \approx 255$ \AA, which is close to $\tilde \lambda  = {\pi  \mathord{\left/
 {\vphantom {\pi  {\left| {\Delta k} \right|}}} \right.
 \kern-\nulldelimiterspace} {\left| {\Delta k} \right|}} = {\pi  \mathord{\left/
 {\vphantom {\pi  {\left| {{k_{\max }} - {k_{\min }}} \right|}}} \right.
 \kern-\nulldelimiterspace} {\left| {{k_{\max }} - {k_{\min }}} \right|}} \approx 471$ a.u.  $ \approx 249$ \AA, where  
 ${k_{\max }} = {\hbar ^{ - 1}}\sqrt {2m{E_{\max }}}  \approx 0.224$ a.u. and ${k_{\min }} = {\hbar ^{ - 1}}\sqrt {2m{E_{\min }}}  \approx 0.217$ a.u.

After pumping stops at  $t > {t_0}$ in case $\Omega  = 0$, the figure (Fig.\ref{fig:FIG7},a) shows that the photoemission pulse of almost rectangular shape almost immediately breaks away from the photoemitter and then moves to the right at a speed close to  ${{\rm{v}}_g}$, experiencing smearing and strong attenuation. In the case of   $\Omega  \ne 0$, at  $t > {t_0}$ a very wide spreading and decaying pulse (Fig.\ref{fig:FIG7}, b) effectively breaks away from the heterostructure not immediately after the pumping stops, but after a time interval of the order of   ${t_0} + {\tau _{R1}}$. It is very important that (Fig.\ref{fig:FIG7},b) (as well as Fig.\ref{fig:FIG5},b) shows spatial oscillations with a wavelength of  $\lambda  = 2260$ \AA, which is close to ${\lambda _{12}} = {{2\pi } \mathord{\left/
 {\vphantom {{2\pi } {\left| {{k_{12}}} \right|}}} \right.
 \kern-\nulldelimiterspace} {\left| {{k_{12}}} \right|}} = {{2\pi } \mathord{\left/
 {\vphantom {{2\pi } {\left| {{{k'}_{R2}} - {{k'}_{R1}}} \right|}}} \right.
 \kern-\nulldelimiterspace} {\left| {{{k'}_{R2}} - {{k'}_{R1}}} \right|}} \approx 2262$ \AA, corresponding to the difference in wave numbers of   ${k'_{R2}} = {\mathop{\rm Re}\nolimits} \left( {{\hbar ^{ - 1}}\sqrt {2m{{E'}_{R2}}} } \right)$= 0.2198 a.u. and   ${k'_{R1}} = {\mathop{\rm Re}\nolimits} \left( {{\hbar ^{ - 1}}\sqrt {2m{{E'}_{R1}}} } \right)$= 0.2184 a.u. For  $t > {t_0}$ and any value of the coordinate  $x$, the time dependence of  $j(x,t)$ has a form similar to Fig.\ref{fig:FIG3},b and Fig.\ref{fig:FIG4},b, demonstrating time oscillations with a difference period ${T_{12}}$, that is  $j(x,t)$ and  $n(x,t)$ exhibits wave properties. The speed of the difference wave of charge and current density is  ${{{{\rm{v}}_{12}} = {\lambda _{12}}} \mathord{\left/
 {\vphantom {{{{\rm{v}}_{12}} = {\lambda _{12}}} {{T_{12}}}}} \right.
 \kern-\nulldelimiterspace} {{T_{12}}}} \approx 4.79 \cdot {10^5}$ m/s. The waves attenuate over a characteristic length of the order of  ${1 \mathord{\left/
 {\vphantom {1 {\left( {\left| {{{k''}_{R1}}} \right| + \left| {{{k''}_{R2}}} \right|} \right)}}} \right.
 \kern-\nulldelimiterspace} {\left( {\left| {{{k''}_{R1}}} \right| + \left| {{{k''}_{R2}}} \right|} \right)}} = 6.94 \cdot {10^4}$a.u.  $ = 3.67 \cdot {10^4}$ \AA. Wave oscillations are present both on the leading edge formed during pumping and on the long ''tail'' formed during the slow decay of quasi-stationary states in the quantum well. This is one of the main results of this paper. Numerical evaluation of the derivatives ${{\partial j(x,t)} \mathord{\left/
 {\vphantom {{\partial j(x,t)} {\partial x}}} \right.
 \kern-\nulldelimiterspace} {\partial x}}$ and  $ - {{\partial n(x,t)} \mathord{\left/
 {\vphantom {{\partial n(x,t)} {\partial t}}} \right.
 \kern-\nulldelimiterspace} {\partial t}}$ by the slope of the envelopes yields values of the same order for them, which agrees with the continuity equation \eqref{eq:math:eq22}. By repeating the pumping pulses at time intervals that are multiples of the periods of the difference oscillations ${T_{12}}$, it is possible to organize a mode of continuous generation of charge density and current waves \cite{bib01,bib02}.

In addition, the figures (Fig.\ref{fig:FIG6}) demonstrate an exponential decrease in the magnitude of the photoemission pulse in the coordinate (and in time, practically according to the law  ${e^{ - {\gamma _p}{{(t - {t_0})} \mathord{\left/
 {\vphantom {{(t - {t_0})} \hbar }} \right. \kern-\nulldelimiterspace} \hbar }}}$). The fact is that the solutions of the kinetic equation for the density matrix used in the calculations were obtained under the assumption of such a strong quantum coherence of electrons that the relaxation processes \textit{inside} the photoemitter determine the type of time dependence of the density matrix, that is, the ''population'' of the corresponding complete system of basis states, despite the fact that for excited states the wave functions are significantly delocalized \textit{outside} the photoemitter. It was already mentioned above in this section that we deliberately chose an overestimated value of the energy dissipation blurring parameter  ${\gamma _p}$ for calculations with a photoemitter in the form of thin metal films, so that it would not differ greatly from the width  $|{\mathop{\rm Im}\nolimits} {E_{R1,2}}|$ of the quasi-stationary levels of the heterostructure, which allowed us to demonstrate the possible competition of the relaxation processes under study on the obtained graphs in the appropriate scale. At a sufficiently small ${\gamma _p}$, the exponential damping will be almost unnoticeable for time intervals  $(t - {t_0}) \ge \hbar /{\gamma _p}$ after the pump is turned off.

However, it is physically obvious that after the formed photoelectron pulse has noticeably separated from the emitter (to the extent that its trailing edge covers the region  $x < {x_3} = 2d$), this pulse loses its strong coherent connection with the emitter and must move ''freely''; it can be considered as a wave packet prepared for the moment the pulse separates from the emitter. The evolution of such a wave packet can be described by \eqref{eq:math:eq12} and \eqref{eq:math:eq13} with a density matrix $\rho _{p',p}^c$ satisfying the simplest differential equation  
$\hbar {{\partial \rho _{p',p}^c} \mathord{\left/
 {\vphantom {{\partial \rho _{p',p}^c} {\partial t}}} \right.
 \kern-\nulldelimiterspace} {\partial t}} = i({E_{p'}} - {E_p})\rho _{p',p}^c$, 
 which has a solution  $\rho _{p',p}^c(t) = \rho _{p',p}^{(2)}({\tilde t_0}){e^{i({E_{p'}} - {E_p}){{(t - {{\tilde t}_0})} \mathord{\left/ {\vphantom {{(t - {{\tilde t}_0})} \hbar }} \right.  \kern-\nulldelimiterspace} \hbar }}}$, where  
 $\tilde{t}_0\geq t_0$, and $\rho _{p',p}^{(2)}({\tilde t_0})$ is given by \eqref{eq:math:eq20} \cite{bib02}. Unlike the standard quantum mechanical wave packet on a pure state \cite{bib01}, such a wave packet cannot be represented as the square of the modulus of a non-stationary wave function, since the integrand  $\rho _{p',p}^{(2)}({\tilde t_0})$ does not split into a product of functions only from  $E$ and only from  $E'$. During its free movement with group velocity  ${{\rm{v}}_g}$, the wave packet spreads out, the wave packet spreads out with a smooth increase in its effective width  $\Delta x(t)$ and a decrease in magnitude  $ \sim {1 \mathord{\left/
 {\vphantom {1 {\Delta x(t)}}} \right.
 \kern-\nulldelimiterspace} {\Delta x(t)}}$ due to dispersion  $E = {{{\hbar ^2}k_3^2} \mathord{\left/
 {\vphantom {{{\hbar ^2}k_3^2} {2m}}} \right.
 \kern-\nulldelimiterspace} {2m}}$ and the difference in phase velocities of its monochromatic components. If the initial width of the prepared wave packet is equal to  $\Delta x$, then the uncertainty relation gives the spread of phase velocities  $\Delta \,{\rm{v  = }}{\hbar  \mathord{\left/
 {\vphantom {\hbar  {m\Delta x}}} \right.
 \kern-\nulldelimiterspace} {m\Delta x}}$, the wave packet becomes twice as wide after a time interval of the order of   $\Delta \,{t_C} \approx 2{{\Delta x} \mathord{\left/
 {\vphantom {{\Delta x} {\Delta \,{\rm{v}}}}} \right.
 \kern-\nulldelimiterspace} {\Delta \,{\rm{v}}}} \approx 2{{m{{(\Delta x)}^2}} \mathord{\left/
 {\vphantom {{m{{(\Delta x)}^2}} \hbar }} \right.
 \kern-\nulldelimiterspace} \hbar }$, and at  $t - {\tilde t_0} \gg \Delta \,{t_C}$ its effective width grows according to the law  $\Delta x(t) \approx {{\hbar (t - {{\tilde t}_0})} \mathord{\left/
 {\vphantom {{\hbar (t - {{\tilde t}_0})} {m\Delta x}}} \right.
 \kern-\nulldelimiterspace} {m\Delta x}}$ .
From Fig.\ref{fig:FIG7} it is clear that when  $\Omega  = 0$ the photoemission pulse is almost completely detached from the emitter already at the moment of  pumping stops  ${\tilde t_0} \approx {t_0} = {10^5}$ a.u. with an initial width of about   $\Delta x \approx {{\rm{v}}_g}{t_0} \approx 1.16 \cdot {10^4}$ \AA (fifth pulse in Fig.\ref{fig:FIG7},a). However, at  $\Omega  = 10$ a.u., due to the slow decay of quasi-stationary states, the pulse does not separate from the emitter for much longer ${\tilde t_0} \approx {t_0} + {\tau _{R2}}\, \approx {10^6}$ a.u. with an initial width of about  $\Delta x \approx {{\rm{v}}_g}({t_0} + {\tau _{R2}}) \approx {10^5}$ \AA  (approximately the ninth pulse in Fig.\ref{fig:FIG7},b).

\section{Dependence on the power of the delta barriers}\label{sec:sec5A}

\begin{figure}[h]
\vspace{-2ex}
\centering
\includegraphics[width=7 cm]{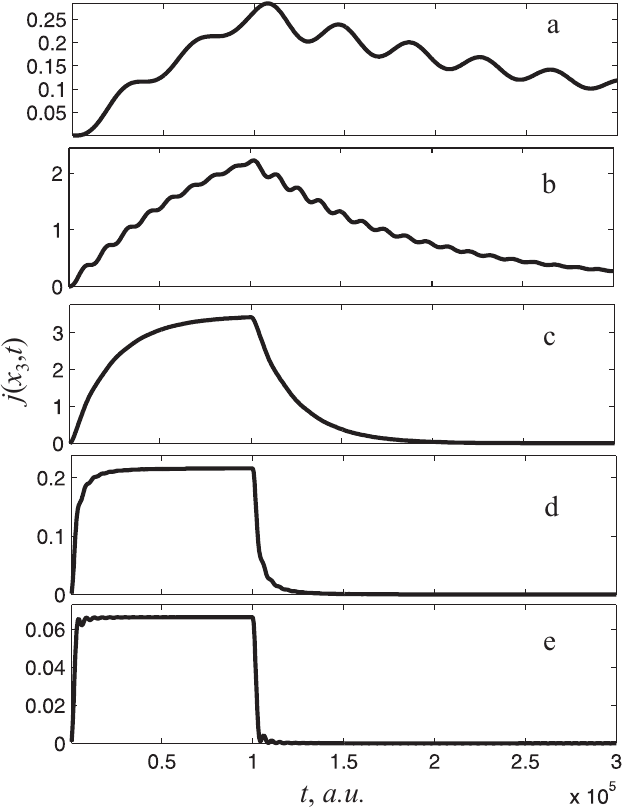}
\caption{. Change in the shape of the photoemission pulse of the current density on the surface  $x = {x_3} = 2d$ of the photoemitter with a decrease in the power  $\Omega $ of the delta barriers: a) $\Omega  = 20$ a.u., (for  $\Omega  = 10$a.u. see (Fig.\ref{fig:FIG3},b), b) $\Omega  = 6$ a.u., ${\rm{c}})\,\Omega  = 2$ a.u.,  ${\rm{d}})\,\Omega  = 0.5$ a.u.,  ${\rm{e}})\,\Omega  = 0$ (the same as (Fig.\ref{fig:FIG3},a). }\label{fig:FIG8}
\end{figure}

The series of figures (Fig.\ref{fig:FIG8}) illustrates the transformation of the photoemission pulse from the quasi-wave form (Fig.\ref{fig:FIG3},b) to the quasi-rectangular form (Fig.\ref{fig:FIG3},a) with a decrease in the power  $\Omega $ of the potential delta barriers.
These graphs show that as  $\Omega $ decreases, the pulse duration monotonically decreases, and its magnitude first increases due to an increase in the transparency of the delta barriers (at the same time, the period of oscillations decreases to zero (Fig.\ref{fig:FIG8},a, Fig.\ref{fig:FIG3},b, Fig.\ref{fig:FIG8}), then the magnitude the pulse decreases without wave oscillations (Fig.\ref{fig:FIG8} c,d,e). This behavior is explained by the evolution of the poles of the partial amplitudes on the complex energy plane: with a decrease in $\Omega $, they shift downwards along the real energy axis with an increase of mutual distances in doublets and upwards along the imaginary axis. Accordingly, the energies of the quasi-stationary levels decrease and their widths increase, which causes a change in their contribution to the integrals over a fixed energy region  ${E_{\min }} \le E,E' \le {E_{\max }}$. It was noted above that the intensity of these contributions can be characterized by the value  ${\left| {{m_r}(E)} \right|^{ - 1}}$, the behavior of which is illustrated by the series of figures (Fig. 9). It is evident that with a decrease in  $\Omega $ the peaks of  ${\left| {{m_r}(E)} \right|^{ - 1}}$ shift to the left with an increase of mutual distances, they become wider, their intensity decreases and at  $\Omega  = 0$ the integration region corresponds to a wide minimum of  ${\left| {{m_r}(E)} \right|^{ - 1}}$.

\begin{figure}[h]
\vspace{-2ex}
\centering
\includegraphics[width=7.25 cm]{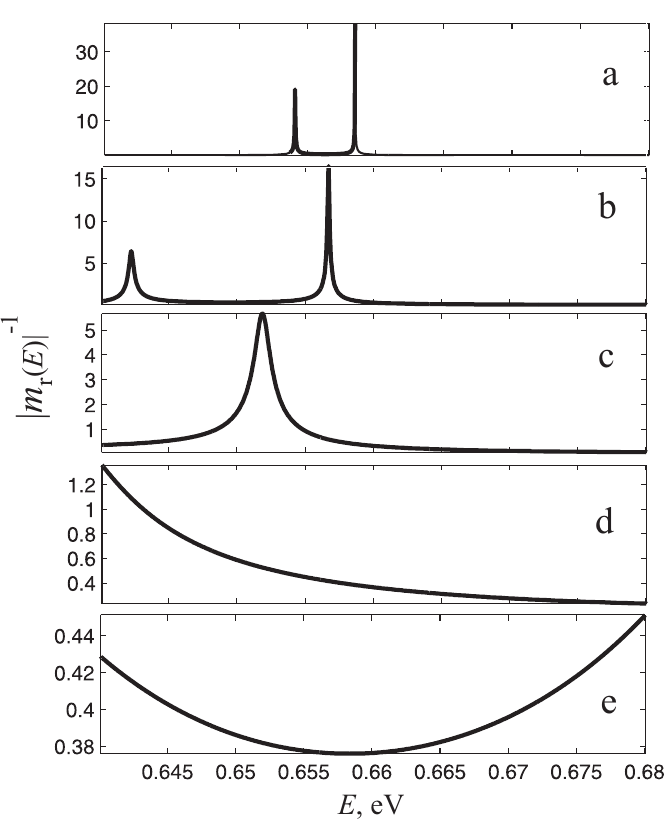}
\caption{Change in dependence  ${\left| {{m_r}(E)} \right|^{ - 1}}$ with a decrease in the power  $\Omega $ of delta barriers:  ${\rm{a}}) \,\Omega  = 20$ a.u., (for  $\Omega  = 10$ a.u. see Fig.\ref{fig:FIG3},b),  ${\rm{b}})\,\Omega  = 6$a.u.  ${\rm{c}})\,\Omega  = 2$ a.u.,  ${\rm{d}})\,\Omega  = 0.5$ a.u.,  ${\rm{e}})\,\Omega  = 0$}\label{fig:FIG9}
\end{figure}

\section{Comparison of the contributions of outgoing and incoming waves}\label{sec:sec6A}

To compare the contributions to the photoemission current of outgoing  ${A_3}$-waves and incoming  ${B_3}$-waves, we calculated using \eqref{eq:math:eq12}-\eqref{eq:math:eq14} these explicit contributions to  $n(x,t)$ and $j(x,t)$, formally setting equal to zero in turn the amplitudes  ${B_3} \equiv B = 0$ and  ${A_3} \equiv A = 0$, which are contained in the factors ${n_{p,p'}}(x)$ and ${j_{p,p'}}(x)$ according \eqref{eq:math:eq3}, \eqref{eq:math:eq4}, and \eqref{eq:math:eq5}. In the case of   ${A_3} \ne 0,{B_3} = 0$, the graphs of the functions  ${n_{B = 0}}(x,t)$ and  ${j_{B = 0}}(x,t)$ have practically the same form as the graphs of $n(x,t)$ and $j(x,t)$, presented in Fig.\ref{fig:FIG3} - Fig.\ref{fig:FIG7} at the same image scale. 

At a large distance from the photoemitter $x \gg {x_3} = 2d$, the differences  $\left| {n(x,t) - {n_{B = 0}}(x,t)} \right| \ll n(x,t)$ and  $\left| {j(x,t) - {j_{B = 0}}(x,(x,t)} \right| \ll j(x,t)$ are relatively small and have a peculiar spatio-temporal structure (as well as the contributions  ${n_{A = 0}}(x,t) \ll n(x,t)$ and  $\left| {{j_{A = 0}}(x,t)} \right| \ll j(x,t)$ of one  ${B_3}$-wave in the case of  ${A_3} = 0,{B_3} \ne 0$), which corresponds to the estimates of  Section 3. Moreover, for the same photoemitter, the dependences on  $t$ and  $x$ of these contributions to  $n(x,t)$ and $j(x,t)$ can differ greatly. The calculated graphs of the time dependence of these quantities at point  $x = {x_4} = 2.5 \cdot {10^4}$ \AA $ \gg {x_3} = 2d$ are shown in the figures (Fig.\ref{fig:FIG10}) - (Fig.\ref{fig:FIG13}):

\begin{figure}[h]
\vspace{-2ex}
\centering
\includegraphics[width=6.45 cm]{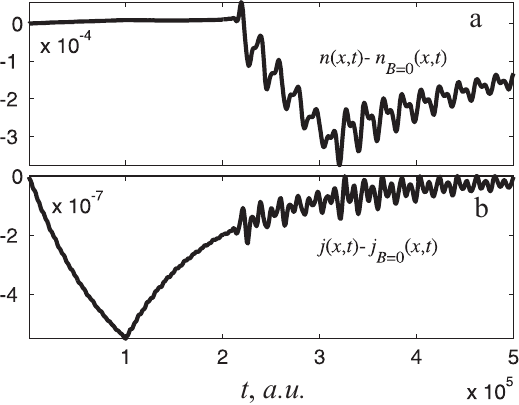}
\caption{Time dependencies a)  $n(x,t) - {n_{B = 0}}(x,t)$ and b)  $j(x,t) - {j_{B = 0}}(x,t)$ at  $\Omega  = 10$a.u. and 
$x = {x_4} = 2.5 \cdot {10^4}$ \AA.}\label{fig:FIG10}
\end{figure}

\begin{figure}[h]
\vspace{-2ex}
\centering
\includegraphics[width=6.45 cm]{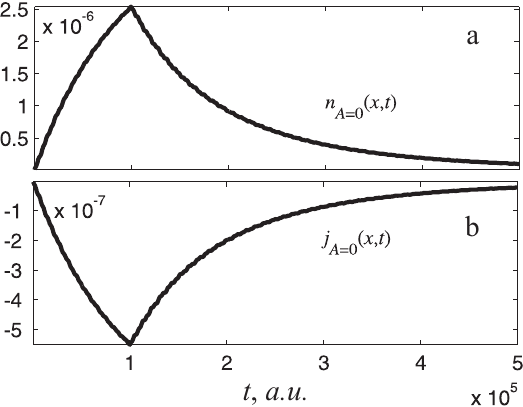}
\caption{Time dependencies a) ${n_{A = 0}}(x,t)$ and b)   ${j_{A = 0}}(x,t)$ at  $\Omega  = 10$ a.u. and $x = {x_4} = 2.5 \cdot {10^4}$ \AA.}\label{fig:FIG11}
\end{figure}

\begin{figure}[h]
\vspace{-2ex}
\centering
\includegraphics[width=6.5 cm]{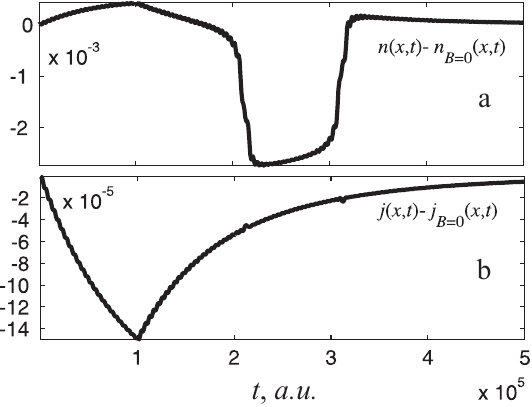}
\caption{Time dependencies a)  $n(x,t) - {n_{B = 0}}(x,t)$ and b)  $j(x,t) - {j_{B = 0}}(x,t)$ at  $\Omega  = 0$ and  $x = {x_4} = 2.5 \cdot {10^4}$ \AA}\label{fig:FIG12}
\end{figure}

\begin{figure}[h]
\vspace{-2ex}
\centering
\includegraphics[width=6.5 cm]{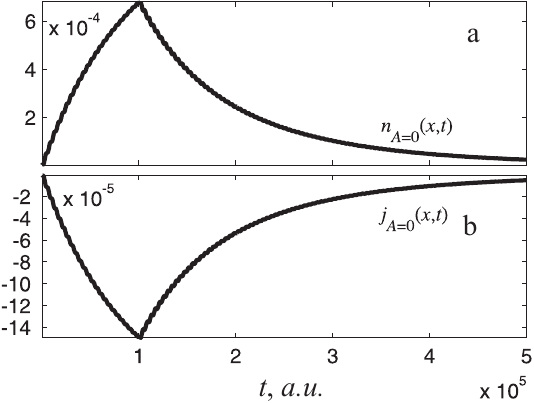}
\caption{Time dependencies a)  ${n_{A = 0}}(x,t)$ and b) ${j_{A = 0}}(x,t)$ at  $\Omega  = 0$ and $x = {x_4} = 2.5 \cdot {10^4}$ \AA.}\label{fig:FIG13}
\end{figure}

All graphs Fig.\ref{fig:FIG10}-Fig.\ref{fig:FIG13} demonstrate the relative smallness of the contributions of incoming  ${B_3}$-waves compared to the contributions of outgoing  ${A_3}$-waves, as well as an anomalous fracture at the moment  $t = {t_0} = {10^5}$ a.u. of the end of the pumping  (except for (Fig.\ref{fig:FIG10},a)), we note that in the scale of Fig.\ref{fig:FIG4}) this fracture does not appear. In the pumping mode for a photoemitter with a heterostructure  $\Omega  = 10$ a.u., the contribution of the  ${B_3}$-waves is   from the contribution of the  ${A_3}$-waves, and for a photoemitter without a heterostructure $\Omega  = 0$, the ratio of these contributions is  $ \sim {10^{ - 4}}$ (recall that here the units of measurement satisfy  $n(x,t) = {\rm{v}}_g^{ - 1}j(x,t) = 4.566\,j(x,t)$). On the graph (Fig.\ref{fig:FIG10}) difference oscillations with the period  ${T_{12}}$ are visible, which appear due to the interference terms  $ \sim {A_3}{B_3}$ in $n(x,t)$ and  $j(x,t)$. On all other graphs, a law of decrease or increase in the time-dependent smooth addition to  $n(x,t)$ and  $j(x,t)$ similar to a hyperbolic law in time  $ \sim {\tilde t^{ - 1}}$ appears, which corresponds to the estimate (\eqref{eq:math:eq27}(a)). 

Calculations similar to (Fig.\ref{fig:FIG10})-(Fig.\ref{fig:FIG13}) for a point on the surface of the photoemitter at $x = {x_3} = 2d$ (as in Fig.\ref{fig:FIG3}) at  $\Omega  = 0$ do not demonstrate the noted smallness of the contributions of incoming  ${B_3}$-waves compared to outgoing  ${A_3}$-waves. These calculations give large magnitudes  $j({x_3},t) - {j_{B = 0}}({x_3},t) \approx  - 0.2$,  ${j_{A = 0}}({x_3},t) \approx  - 0.2$, but in such a way that the total current remains small $j({x_3},t) \approx 0.06$ (Fig.\ref{fig:FIG3},a). This demonstrates the strong influence of the impermeable wall at  $x = 0$ with the formation of a standing wave at  $x = {x_3} = 2d$. However, for a photoemitter with delta barriers   $\Omega  = 10$ a.u. the indicated magnitudes are small  $j({x_3},t) - {j_{B = 0}}({x_3},t) \approx  - 2 \cdot {10^{ - 4}}$, ${j_{A = 0}}({x_3},t) \approx  - 2 \cdot {10^{ - 4}}$, while the total current is not small  $j({x_3},t) = 1$ (Fig.\ref{fig:FIG3},b).

\begin{figure}[h]
\vspace{-2ex}
\centering
\includegraphics[width=6.45 cm]{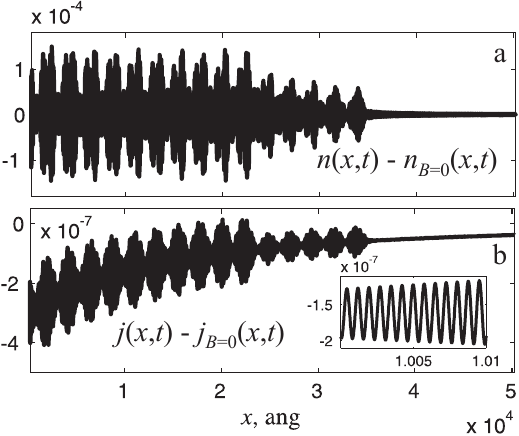}
\caption{Coordinate dependencies a)  $n(x,t) - {n_{B = 0}}(x,t)$ and b) $j(x,t) - {j_{B = 0}}(x,t)$ at  $t = 3 \cdot {10^5}$ a.u.,  $\Omega  = 10$ a.u.}\label{fig:FIG14}
\end{figure}

\begin{figure}[h]
\vspace{-2ex}
\centering
\includegraphics[width=6.5 cm]{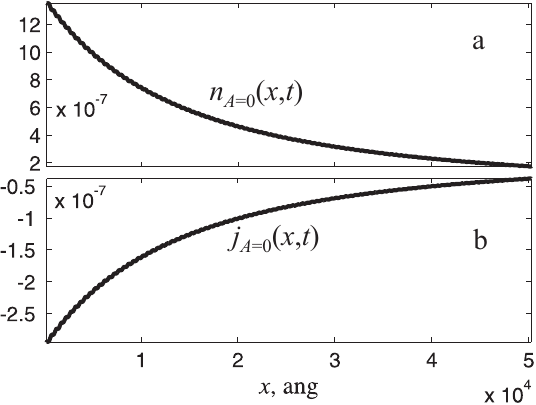}
\caption{Coordinate dependencies a)  ${n_{A = 0}}(x,t)$ and b)  ${j_{A = 0}}(x,t)$ at  $t = 3 \cdot {10^5}$ a.u.,  $\Omega  = 10$ a.u.    }\label{fig:FIG15}
\end{figure}

\begin{figure}[h]
\vspace{-2ex}
\centering
\includegraphics[width=6.6 cm]{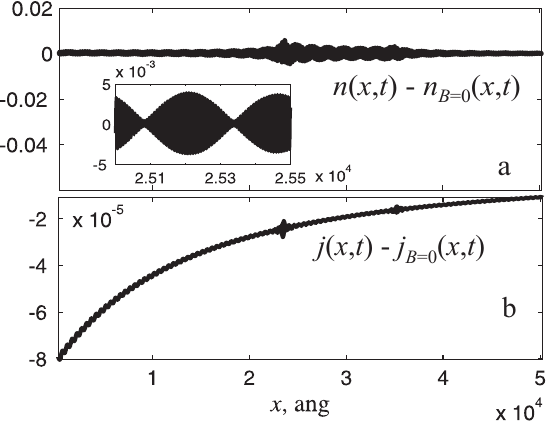}
\caption{Coordinate dependencies a)  $n(x,t) - {n_{B = 0}}(x,t)$and b)  $j(x,t) - {j_{B = 0}}(x,t)$ at  $ at = 3 \cdot {10^5}$ a.u.,  $\Omega  = 0$}\label{fig:FIG16}
\end{figure}

\begin{figure}[h]
\vspace{-2ex}
\centering
\includegraphics[width=6.35 cm]{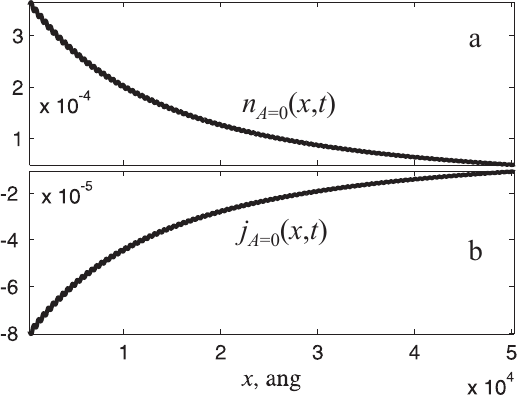}
\caption{Coordinate dependencies a) $n_{A = 0}(x,t)$  and b) ${j_{A = 0}}(x,t)$ at $t= 3 \cdot {10^5}$ a.u.,  $\Omega  = 0$      }\label{fig:FIG17}
\end{figure}

The figures Fig.\ref{fig:FIG14} - \ref{fig:FIG17} show the calculated graphs of the spatial dependence of the same quantities to the right of the heterostructure  $x \ge {x_3} = 2d$ at the moment of time  $t = 3 \cdot {10^5}$a.u. $ > {t_0}$ for a pulse intermediate between the sixth and seventh pulses in Fig.\ref{fig:FIG7}.

The graphs (Fig.\ref{fig:FIG15}) and Fig.\ref{fig:FIG17}) show that in the photoelectron pulse propagation mode the contributions of the incoming  ${B_3}$-waves  ${n_{A = 0}}(x,t)$ and ${j_{A = 0}}(x,t)$ are smooth decreasing functions of the  $x$-coordinate and make up a $10^{-4}-10^{-7}$  of the contributions of the outgoing  ${A_3}$-waves. For a photoemitter with a heterostructure at  $\Omega  = 10$ a.u., due to the interference terms  $ \sim {A_3}{B_3}$, the coordinate dependences $j(x,t) - {j_{B = 0}}(x,t)$ and  $n(x,t) - {n_{B = 0}}(x,t)$ oscillate in space (Fig.\ref{fig:FIG14}) with a carrier wavelength  $\lambda  = \pi /{k_g} \approx 7.6$ \AA and with an envelope in the form of a standing wave with a wavelength  $\lambda  = 2270$ \AA, which is close to the resonant difference wavelength  ${\lambda _{12}} = {{2\pi } \mathord{\left/
 {\vphantom {{2\pi } {\left| {{k_{12}}} \right|}}} \right.
 \kern-\nulldelimiterspace} {\left| {{k_{12}}} \right|}} = {{2\pi } \mathord{\left/
 {\vphantom {{2\pi } {\left| {{{k'}_{R2}} - {{k'}_{R1}}} \right|}}} \right.
 \kern-\nulldelimiterspace} {\left| {{{k'}_{R2}} - {{k'}_{R1}}} \right|}} \approx 2280$\AA. In the absence of a heterostructure at $\Omega  = 0$, the interference contribution  $ \sim {A_3}{B_3}$ to the current density  $j(x,t) - {j_{B = 0}}(x,t)$ decreases quite smoothly with increasing  $x$ (Fig.\ref{fig:FIG16},b), but this contribution to the charge density $n(x,t) - {n_{B = 0}}(x,t)$ (Fig.\ref{fig:FIG16},a) noticeably oscillates in space with a carrier wavelength  $\lambda  = \pi /{k_g} \approx 7.6$ \AA and with an envelope in the form of a standing wave with a wavelength  $\lambda  = 1700$ \AA.

\section{Transition to stationary mode}\label{sec:sec7A}

\begin{figure}[h]
\vspace{-2ex}
\centering
\includegraphics[width=7.1 cm]{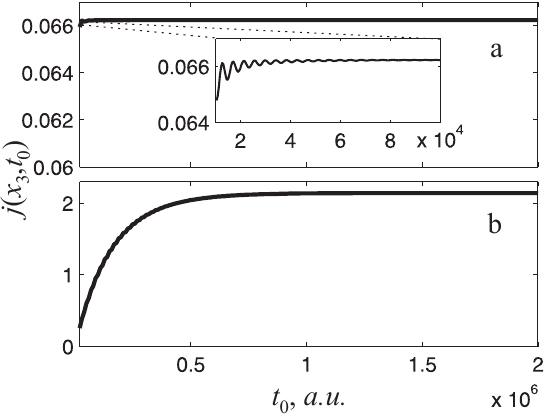}
\caption{Photocurrent density $j({x_3},{t_0})$  at the boundary  $x = {x_3} = 2d$ depending on the pumping time a) for $\Omega  = 0$ a.u., b) for   $\Omega  = 10$ a.u. }\label{fig:FIG18}
\end{figure}

Calculations of the photocurrent density in the pumping mode show that with a change in the pumping time $t_0$   in the absence of delta barriers $\Omega  = 0$, the current density  $j({x_3},{t_0})$ at the boundary  $x = {x_3} = 2d$ oscillates with a period of  $\Delta {t_0} \approx 4.3 \cdot {10^3}$ a.u.  $ \approx 1.04 \cdot {10^{ - 13}}$ s, while the product  ${{\rm{v}}_g}\Delta {t_0} \approx 0.219 \cdot 4.3 \cdot {10^3}$ a.u.  $ \approx 942$ \AA coincides with the wavelength  $\lambda  = {{2\pi } \mathord{\left/
 {\vphantom {{2\pi } {\left| {\Delta k} \right|}}} \right.
 \kern-\nulldelimiterspace} {\left| {\Delta k} \right|}} = {{2\pi } \mathord{\left/
 {\vphantom {{2\pi } {\left| {{k_{\max }} - {k_{\min }}} \right|}}} \right.
 \kern-\nulldelimiterspace} {\left| {{k_{\max }} - {k_{\min }}} \right|}} \approx 941$ a.u. corresponding to the spectral width of the detected electron pulse, and with an increase in the pumping time  ${t_0}$, the value  $j({x_3},{t_0})$ increases on average and at  ${t_0} \to \infty $ tends to a constant value  $j({x_3},\infty ) \approx 0.066$ (Fig.\ref{fig:FIG18},a). In the presence of delta barriers  $\Omega  = 10$a.u, the current density at the boundary  $j({x_3},{t_0})$ smoothly increases with  ${t_0}$ and at  ${t_0} \to \infty $ tends to a value  $j({x_3},\infty ) \approx 2.14$  (Fig.\ref{fig:FIG18},b). The values  $j({x_3},\infty )$ are equal to the current densities of steady-state pumping at the boundary. 

\begin{figure}[h]
\vspace{-2ex}
\centering
\includegraphics[width=7.9 cm]{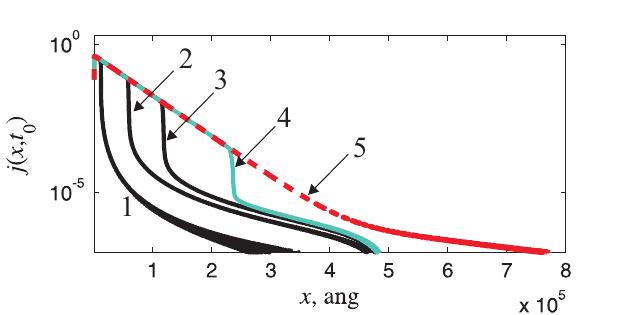}
\caption{For  $\Omega  = 10$ a.u.  the coordinate dependence of the current density  $j(x,{t_0})$ at $x > {x_3} = 2d$ in the moment of pumping termination  $t = {t_0}$ at: 1)  ${t_0} = {10^5}$ a.u. (the same as \eqref{eq:math:eq5} in Fig.\ref{fig:FIG5},a), 2)  ${t_0} = 5 \cdot {10^5}$a.u., 3)  ${t_0} = {10^6}$a.u., 4)  ${t_0} = 5 \cdot {10^6}$a.u., 5)  ${t_0} = \infty $ a.u.}\label{fig:FIG19}
\end{figure}

\begin{figure}[h]
\vspace{-2ex}
\centering
\includegraphics[width=6.75 cm]{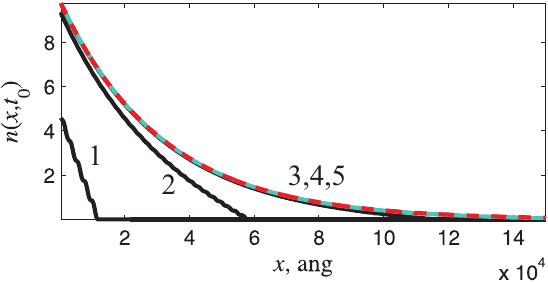}
\caption{For  $\Omega  = 0$ the coordinate dependence of the current density  $j(x,{t_0})$ at  $x > {x_3} = 2d$  in the moment of pumping termination  $t = {t_0}$ at: 1)  ${t_0} = {10^5}$ a.u. (the same as \eqref{eq:math:eq5}  in Fig.\ref{fig:FIG5},b), 2)  ${t_0} = 5 \cdot {10^5}$ a.u., 3)  ${t_0} = {10^6}$ a.u., 4)  ${t_0} = 5 \cdot {10^6}$ a.u., 5)  ${t_0} = \infty $ a.u.}\label{fig:FIG20}
\end{figure}

With increasing time  $t < {t_0}$, the right pulse thresholds in the figures Fig.\ref{fig:FIG5} move to the right with a smooth decrease in their slope. With increasing pumping duration  $t_0$, the current density at any point  $x > {x_3} = 2d$ smoothly increases with increasing  ${t_0}$ and at  ${t_0} \to \infty $ asymptotically tends to the maximum for this point current density  $j(x,\infty )$ of the steady-state pumping . However, the value of the  $x$-component of the current density at the moment of pumping termination  $j(x,{t_0})$ and, in particular, the magnitude  $j(x,\infty)$ of the steady-state pumping current with increasing distance from the photoemitter $x$  smoothly decrease (Fig.\ref{fig:FIG19}) and (Fig.\ref{fig:FIG20}).

This behavior of the photocurrent is the result of several competing processes of different rates: 1) pumping and dissipation leaks inside the photoemitter, 2) the exit of electrons from the photoemitter through potential barriers, 3) the spreading of the wave pulse (the resulting wave packet) due to the difference in phase velocities of the monochromatic components and the movement forward the fastest of them. 

Note also that in the case under consideration the photocurrent detector is located in a plane at some finite distance   at vacuum potential and does not create an electron-extracting electric field, so we perform calculations in the basis of exponential traveling plane waves. To obtain a non-zero stationary current at  $x \to \infty $ it is necessary to apply an additional positive potential to the detector and perform calculations in the basis of the Airy function type.

Developing calculations similar to the fifth curve in Fig.\ref{fig:FIG5}, we obtained the coordinate dependence curves of the current density  $j(x,{t_0})$ at the moment of pumping stop   $t = {t_0}$ at $x > {x_3} = 2d$. Fig.\ref{fig:FIG19} shows this dependence for a thin-film photoemitter without delta barriers  $\Omega  = 0$ in a semi-logarithmic scale, since in this case the current decreases very quickly with distance. The leading edge shows regions of different slopes: the base of the leading edge (leader) moves forward faster than its rear part, which gradually disappears, asymptotically approaching the curve 5) of the steady-state pumping current density $j(x,\infty)$.

In Fig.\ref{fig:FIG20} this dependence  $j(x,{t_0})$ is shown for a film photoemitter with two delta barriers  $\Omega  = 10$ a.u. In this case the current decreases more slowly with distance. The inclined front is rather quickly absorbed by the leader between curves 2)  ${t_0} = 5 \cdot {10^5}$ a.u. and 3)  ${t_0} = {10^6}$ a.u., which then smoothly passes into curve 5) of the steady-state pumping current density  $j(x,\infty)$.

\section{Conclusions}\label{sec:sec8A}

Thus, in the process of fast pulsed photoemission from a flat thin-film photoemitter formed by a double quantum well on a dielectric substrate, it is possible to generate charge and current density waves, which arise as a result of the population and slow decay of quasi-stationary states of the doublet of the double quantum well during photoexcitation of electrons directly in the conducting layers from inside this well. To describe this phenomenon, we considered a quasi-one-dimensional model of the system, within the framework of which it is obvious that there is no alternative to using the strict quantum theory of the atomic photoeffect and the scenario of photoemission from a crystal as an inverse LEED process \cite{bib23}-\cite{bib31}. With such a description, the main goal is usually the calculation of time-independent averaged probability and effective cross section of the atomic photoeffect, or the density of the stationary current of photoemission from crystals. In this paper, we have implemented this algorithm for the first time to calculate the explicit spatio-temporal dependence of pulsed charge and current densities.

Outside the photoemitter, there is a plane-wave exponential that plays a major role, describing a wave propagating toward the detector, and its partial amplitude is determined by normalization and depends weakly on energy. Partial amplitudes of incoming waves, as well as waves inside the heterostructure, are expressed through it using boundary conditions of the time-reversed scattering problem; in the presence of a double-well heterostructure, they acquire pole singularities similar to the singularities of the amplitudes of direct electron scattering on this heterostructure with an accuracy of complex conjugation. The same pole singularities are acquired by the matrix elements of the electron dipole moment.

       For fast pulse processes, the charge and current densities are given by sums over the corresponding stationary states of the products of the time-dependent elements of the density matrix and the coordinate-dependent matrix elements of the charge and current densities. The density matrix elements satisfy the kinetic equation of quantum optics. We obtained solutions to this equation for cases of abrupt switching on and off of the pumping light pulse. Then we calculated and estimated the integral sums expressing both the total charge and current densities of photoemission in a given energy range recorded by the detector and the partial contributions of outgoing and incoming waves. For a flat photoemitter with a double-well heterostructure, the wave oscillations of the current were manifested both at the leading edge of the pulse formed during the pumping process and at the long "tail" formed during the slow decay of quasi-stationary states. By repeating the pumping pulses at time intervals multiple of the periods of the difference oscillations, it is possible to organize a mode of continuous generation of charge and current density waves in the picosecond range.
       
        For comparison, calculations were also made for photoemission from a metal film on a substrate without a heterostructure. The dependence of the photocurrent on the power of the barriers of the heterostructure was investigated. It was shown that the double-well heterostructure increases, stabilizes and stretches the photocurrent pulse. 
Asymptotic analytical estimates were made in terms of wave packet evolution using the extreme phase and fastest descent methods. The contributions of the outgoing and incoming waves from the detector side are compared, and the relative smallness of the incoming wave contribution is demonstrated.  The limiting transition to the stationary pumping regime was numerically analyzed.  In the next article, we will present the results of a study of fast pulsed photoemission from a flat thin-film photoemitter formed by a double quantum well without a dielectric substrate, when incoming electron waves are present on both sides of the emitter.

\bmhead{Acknowledgements}

The work (A.A.)  is partially supported by the Ministry of Science and Higher Education of Russian Federation under the project FSUN-2023-0006.


\begin{thebibliography}{9}

\bibitem{bib01} Yu. G. Peisakhovich and A.A. Shtygashev, Formation of probability and current waves at the scattering of a Gaussian wave packet by a double quantum well. J. Phys. A: Math. Theor. \textbf{56}, 115302, (2023).

\bibitem{bib02}	Yu. G. Peisakhovich and A.A. Shtygashev, Density matrix theory of the charge and current wave formation at fast pulsed photoemission through a double quantum well.  Phys. Rev. B \textbf{106}, 115405, (2022).

\bibitem{bib03}	A. del Campo, G. Garcia-Calderon, and J. Muga, Quantum transients. Phys. Rep. \textbf{476}, 1 (2009)

\bibitem{bib04}	K. Leo at al., Coherent oscillations of a wave packet in a semiconductor double-quantum-well structure. Phys. Rev. Lett. \textbf{66}, 201(1991)

\bibitem{bib05}	H. G. Roskos at al., Coherent submillimeter-wave emission from charge oscillations in a double-well potential. Phys. Rev. Lett. \textbf{68}, 2216 (1992)

\bibitem{bib06}	R. Romo, J. Villavicencio, and G. Garcia-Calderon, Transient tunneling effects of resonance doublets in triple barrier systems. Phys. Rev. B \textbf{66}, 033108 (2002)

\bibitem{bib07}	Yu. G. Peisakhovich and A.A. Shtygashev, Formation of a quasistationary state by scattering of wave packets on a finite lattice. Phys. Rev. B \textbf{77}, 075326 (2008)

\bibitem{bib08}	Yu. G. Peisakhovich and A.A. Shtygashev, Formation of a quasistationary state by Gaussian wave packet scattering on a lattice of ?? identical delta potentials.  Phys. Rev. B \textbf{77}, 075327 (2008)

\bibitem{bib09}	 G. Garcia-Calderon, R. Romo, and J. Villavicencio, Internal dynamics of multibarrier systems for pulsed quantum decay.  Phys. Rev.  A \textbf{79}, 052121 (2009).

\bibitem{bib10} 	S. Cordero, G. Garcia-Calderon, R. Romo, and J. Villavicencio, Unified analytical description of the time evolution of decay for initial states formed by wave-packet scattering and by initial decaying states in quantum systems.  Phys. Rev. A  \textbf{ 84}, 042118 (2011).

\bibitem{bib11} Yu. A. Ilyinsky and L.V. Keldysh, Interaction of electromagnetic radiation with matter, (Moscow State Univ. Publ. House, Moscow, 1989, in Russian).

\bibitem{bib12}     A.I. Kopeliovich, Effect of Electron Collisions on Interband Transitions in Metals. Zh. Eksp. Teor. Fiz. \textbf{58} 601 (1970) [ Sov.Phys.-JETP. \textbf{31}, 323 (1970)].

\bibitem{bib13}	V.M. Nabutovskii and Yu. G. Peisakhovich, Singularities in the energy distribution of photoelectrons.  Zh. Eksp. Teor. Fiz. \textbf{70}, 1081 (1976) [ Sov. Phys.-JETP. \textbf{43}, 564 (1976)].

\bibitem{bib14} T. Hertel, E. Knoesel, W.Wolf, and G. Ertl, Ultrafast Electron Dynamics at Cu(111): Response of an Electron Gas to Optical Excitation.  Phys. Rev. Lett.  \textbf{76}, 535, (1996).

\bibitem{bib15}	M. Bauer, S. Pawlik, and M. Aeschlimann, Decay dynamics of photoexcited alkali chemisorbates:?Real-time investigations in the femtosecond regime. Phys. Rev.  B \textbf{60}, 5016 (1999).

\bibitem{bib16} T. Klamroth, P. Saalfrank, and U. Hofer, Open-system density-matrix approach to image-potential dynamics of electrons at Cu(100):?Energy- and time-resolved two-photon photoemission spectra.  Phys. Rev. B \textbf{64}, 035420 (2001)

\bibitem{bib17} M. Weiner, Femtosecond pulse shaping using spatial light modulators. Rev. Sci. Instrum. \textbf{71}, 1929, (2000).

\bibitem{bib18} 	A.  Abbaszadeh, A. Tehranian, and J.A. Saleni, Phase-only femtosecond optical pulse shaping based on an all-dielectric polarization-insensitive metasurface. Opt. Express. \textbf{29}, 36900 (2021).

\bibitem{bib19} Y. Park, M.H. Asghari, T.-J. Ahn, and J. Azana, Transform-limited picosecond pulse shaping based on temporal coherence synthesization. 	Opt. Express. \textbf{15}, 9584 (2007).

\bibitem{bib20}	 I. Will and G. Klemz, Generation of flat-top picosecond pulses by coherent pulse stacking in a multicrystal birefringent filter. Opt. Express. \textbf{16}, 14922 (2008).

\bibitem{bib21}	 P. Petropoulos, M. Ibsen, A.D. Ellis, and D.J. Richardson, Rectangular pulse generation based on pulse reshaping using a superstructured fiber Bragg grating. J. Lightwave Technol. \textbf{19}, 746 (2001).

\bibitem{bib22}	 A. Pakhomov, N. Rosanov, M. Arkhipov, and R. Arkhipov, Sub-10 fs unipolar pulses of a tailored waveshape from a multilevel resonant medium.  Opt. lett. \textbf{48}, 6504 (2023).

\bibitem{bib23}	 M. Stobbe, Zur Quantenmechanik photoelektrischer Prozesse. Annalen der Physik, \textbf{399}, 661, (1930). 

\bibitem{bib24}	 V.B. Berestetskii, E. M. Lifshitz, and L.P. Pitaevskii, Quantum electrodynamics. Butterworth-Heinemann (1982).

\bibitem{bib25}	 M. Yu. Amusia, Atomic photoelectric effect. Publishing house ''Nauka'', Moscow, in Russian (1987).

\bibitem{bib26}	S. Huffner, Photoelectron Spectroscopy. Principles and Applications. (Springer, 2010).

\bibitem{bib27}	Solid-State Photoemission and Related Methods. Theory and Experiment. Eds. W. Schattke and M.A.Van Hove, (Wiley-VCH, 2003). 

\bibitem{bib28} P.J. Feibelman and D.E. Eastman, Photoemission spectroscopy -Correspondence between quantum theory and experimental phenomenology. Phys. Rev. B \textbf{10}, 4932 (1974).

\bibitem{bib29}	 I. Adawi, Theory of the Surface Photoelectric Effect for One and Two Photons.  Phys. Rev. \textbf{134}, A788 (1964).

\bibitem{bib30}	 R.E.B. Makinson, The Surface Photoelectric Effect. Phys. Rev. \textbf{75}, 1908 (1949).

\bibitem{bib31}	 C. Caroli, D. Lederer-Rozenblatt, B. Roulet, and D. Saint-James, Inelastic Effects in Photoemission: Microscopic Formulation and Qualitative Discussion.  Phys. Rev. B \textbf{8}, 4552 (1973).

\bibitem{bib32}	 W.E. Spicer, Photoemissive, Photoconductive, and Optical Absorption Studies of Alkali-Antimony Compounds.  Phys. Rev. \textbf{112}, 114 (1958).

\bibitem{bib33} W.E. Spicer and A. Herrera-Gomes, Modern Theory and Applications of Photoemitters. in Proceedings of Spie - the International Society For Optical Engineering. 2022: 18-35 (San Diego, 1993), SLAC-PUB-6306 and SLAC/SSRL-0042 (Aug 1993)
\bibitem{bib34}	 J. C. Phillips. The Fundamental Optical Spectra of Solids. Solid State Physics, \textbf{18}, (1966).
\bibitem{bib35} F. Bassani and G. Pastori Parravicini, Electronic States and Optical Transitions in Solids. Pergamon Press, (1975).

\bibitem{bib36}	 L. D. Landau and E. M. Lifshitz, Quantum Mechanics. Non-Relativistic Theory. (Oxford: Pergamon Press) (1977).

\bibitem{bib37}Yu. G. Peisakhovich, The recurrent algorithm of the rigorous solving 1-dimensional wave equations in multilayered media. J. Phys. A: Math. and Gen., \textbf{29}, 5103 (1996)
\bibitem{bib38}	Yu. G. Peisakhovich, The transfer matrices and electronic spectrum of the multilayered system in a homogeneous magnetic field. J. Phys. A: Math. and Gen., \textbf{32}, 3133 (1999)


\end{thebibliography}
\end{document}